\documentclass[fleqn,usenatbib]{mnras}

\usepackage{newtxtext,newtxmath}
\usepackage[T1]{fontenc}

\DeclareRobustCommand{\VAN}[3]{#2}
\let\VANthebibliography\thebibliography
\def\thebibliography{\DeclareRobustCommand{\VAN}[3]{##3}\VANthebibliography}

\usepackage{graphicx}	% Including figure files
\usepackage{amsmath}	% Advanced maths commands
\usepackage{color,soul} % highlighting (temp)
\usepackage{verbatim}   % Comments (temp)

\usepackage{tikz,xcolor,hyperref}

\newcommand{\mgii}{Mg\,\textsc{ii}}
\newcommand{\civ}{C\,\textsc{iv}}

\newcommand{\hbeta}{H\textsc{$\beta$}}

\newcommand{\mgiil}{Mg\,\textsc{ii}$\lambda$2799}

\defcitealias{Onken_2022_QLF}{O22}
\defcitealias{Trakhtenbrot_2011}{T11}
\defcitealias{He_2024_H24}{H24}
\defcitealias{Lopez_2016_XQ100}{XQ100}
\defcitealias{Lai_XQz5}{XQz5}
\defcitealias{Wu_2022_Demo}{Wu et al. 2022}
\defcitealias{Li_2023}{Li et al. 2023}
\defcitealias{Schulze_2015}{Schulze et al. 2015}
\defcitealias{Ananna_2022}{Ananna et al. 2022}

\definecolor{lime}{HTML}{A6CE39}
\DeclareRobustCommand{\orcidicon}{%
    \begin{tikzpicture}
    \draw[lime, fill=lime] (0,0) 
    circle [radius=0.16] 
    node[white] {{\fontfamily{qag}\selectfont \tiny ID}};
    \draw[white, fill=white] (-0.0625,0.095) 
    circle [radius=0.007];
    \end{tikzpicture}
    \hspace{-2mm}
}

\newcommand{\orcidChrisO}{\href{https://orcid.org/0000-0003-0017-349X}{\orcidicon}}
\newcommand{\orcidChrisW}{\href{https://orcid.org/0000-0002-4569-016X}{\orcidicon}}
\newcommand{\orcidSamuel}{\href{https://orcid.org/0000-0001-9372-4611}{\orcidicon}}
\newcommand{\orcidFuyan}{\href{https://orcid.org/0000-0002-1620-0897}{\orcidicon}}
\newcommand{\orcidfanxiaohui}{\href{https://orcid.org/0000-0003-3310-0131}{\orcidicon}}

\title[XQz5+ Black Hole Mass Function]{Supermassive black holes are growing slowly by z\(\sim \)5}

\author[S. Lai et al.]{Samuel Lai,$^{1,2}$\orcidSamuel\thanks{E-mail: samuel.lai@anu.edu.au}
Christopher A. Onken,$^{1,3}$\orcidChrisO\,
Christian Wolf,$^{1,3}$\orcidChrisW\,
Fuyan Bian,$^{4}$\orcidFuyan\,
and Xiaohui Fan$^{5}$\orcidfanxiaohui
\\
$^{1}$Research School of Astronomy and Astrophysics, Australian National University, Canberra, ACT 2611, Australia\\
$^{2}$Commonwealth Scientific and Industrial Research Organisation (CSIRO), Space \& Astronomy, P. O. Box 1130, Bentley, WA 6102, Australia\\
$^{3}$Centre for Gravitational Astrophysics, Research Schools of Physics, and Astronomy and Astrophysics, Australian National University, Canberra, ACT 2611, Australia\\
$^{4}$European Southern Observatory, Alonso de C\'{o}rdova 3107, Casilla 19001, Vitacura, Santiago 19, Chile\\
$^{5}$Steward Observatory, University of Arizona, 933 N Cherry Ave, Tucson, AZ 85721, USA\\
}

\date{Accepted XXX. Received YYY; in original form ZZZ}

\pubyear{2023}

\begin{document}
\label{firstpage}
\pagerange{\pageref{firstpage}--\pageref{lastpage}}
\maketitle

\begin{abstract}
We investigate the black hole mass function at $z\sim5$ using XQz5, our recent sample of the most luminous quasars between the redshifts $4.5 < z < 5.3$. We include 72 quasars with black hole masses estimated from velocity-broadened emission-line measurements and single-epoch virial prescriptions in the footprint of a highly complete parent survey. The sample mean Eddington ratio and standard deviation is $\log\lambda \approx -0.20\pm0.24$. The completeness-corrected mass function is modelled as a double power-law, and we constrain its evolution across redshift assuming accretion-dominated mass growth. We estimate the evolution of the mass function from $z=5-4$, presenting joint constraints on accretion properties through a measured dimensionless e-folding parameter, $k_{\rm{ef}} \equiv \langle\lambda\rangle U (1-\epsilon)/\epsilon = 1.79\pm0.06$, where $\langle\lambda\rangle$ is the mean Eddington ratio, $U$ is the duty cycle, and $\epsilon$ is the radiative efficiency. If these supermassive black holes were to form from seeds smaller than $10^8\,M_{\odot}$, the growth rate must have been considerably faster at $z\gg5$ than observed from $z=5-4$. A growth rate exceeding $3\times$ the observed rate would reduce the initial heavy seed mass to $10^{5-6}\,M_{\odot}$, aligning with supermassive star and/or direct collapse seed masses. Stellar mass ($10^2\,M_{\odot}$) black hole seeds would require $\gtrsim4.5\times$ the observed growth rate at $z\gg5$ to reproduce the measured active black hole mass function. A possible pathway to produce the most extreme quasars is radiatively inefficient accretion flow, suggesting black holes with low angular momentum or photon trapping in supercritically accreting thick discs.
\end{abstract}


%%%%%%%%%%%%%%%%%%%%%%%%%%%%%%%%%%%%%%%%%%%%%%%%%%

%%%%%%%%%%%%%%%%% BODY OF PAPER %%%%%%%%%%%%%%%%%%

\section{Introduction} \label{sec:Introduction}
The radiative output of accreting supermassive black holes (SMBHs), which reside in the centres of massive galaxies, is powered by gravitational energy associated with infalling material within the central potential well \citep[e.g.][]{Rees_1984}, allowing them to outshine their hosts by orders of magnitude. The extreme luminosities reached by these systems enable them to be observed across cosmological distances, up to $z \sim 7.5$ \citep[e.g.,][]{Banados_2018, Yang_2020_z7.5, Wang_2021_z7.642}, when the age of the Universe was less than 700 Myr. It is now widely believed that these SMBHs play an important role in regulating host galaxy evolution \citep[see review by][]{Kormendy_2013_Review}, as evidenced by strong correlations between SMBH mass and velocity dispersion \citep[e.g.][]{Magorrian_1998, Ferrarese_2000, Gebhardt2000, Merritt_2001, Kormendy_2011Natur.469..374K}, bulge mass \citep[e.g.][]{Kormendy_Richstone_1995, Marconi_2003, Haring_2004, Peng_2006, Greene_2010} or total stellar mass \cite[e.g.][]{Cisternas_2011, Reines_2015, Davis_2019, Ding_2020, Smethurst_2023}. These tight relations leave little doubt of the existence of feedback mechanisms controlling the host galaxy and SMBH co-evolution \citep[see review by][]{Fabian_2012}. Understanding this co-evolution would aid in the development of a coherent model of galaxy formation and evolution. 

In the study of host galaxy and SMBH co-evolution, the black hole mass is an important parameter that is correlated with other galactic properties. However, the best estimates of SMBH mass suggest that the most massive of them are $\gtrsim 10^9 M_{\odot}$ even at early cosmic times, raising questions about the SMBH cosmic mass assembly and the size of black hole seeds \citep[see review by][]{Volonteri_2021_review}. Accretion-dominated mass assembly implies exponential growth with an e-folding time of $450\lambda^{-1}\epsilon(1-\epsilon)^{-1}$ Myr, where $\epsilon$ is the radiative efficiency and $\lambda$ is the Eddington ratio, $L_{\rm{bol}}/L_{\rm{edd}}$, with $L_{\rm{edd}} = 1.26\times10^{38}\,M_{\rm{BH}}/M_{\odot}\,\rm{erg\, s^{-1}}$. In order to grow $10^9 M_{\odot}$ black holes within 1 Gyr through accretion, $> 10^2 \,M_{\odot}$ black hole seeds would be required by 300 Myr ($z\sim14$), assuming uninterrupted Eddington accretion with the fiducial $\epsilon = 0.1$ radiative efficiency based on the So\l{}tan argument \citep{Soltan_1982_argument, Yu_2002}. Other than accretion, black hole mergers also contribute to the evolution of black hole mass. Large-scale hydrodynamical simulations suggest that the role of mergers in the total SMBH mass budget is secondary to growth by accretion \citep[e.g.][]{Dubois_2014, Kulier_2015, Martin_2018}. However, at late times when galaxies are comparatively gas-poor, the merger contribution to the total mass budget can increase, especially for the highest mass black holes \citep[e.g.][]{Shankar_2013_continuity, Dubois_2014, Kulier_2015}. The role of mergers at high-redshift may be illuminated by the next-generation of gravitational wave interferometers which will extend the redshift horizon and significantly improve the sensitivity to probe mergers of intermediate-mass black holes \citep{Punturo_2010_Einstein, Amaro-Seoane_2017, Reitze_2019}. 

In order to explain the observed high-redshift black hole masses, SMBHs must have formed and grown rapidly \citep[e.g.][]{Volonteri_2012}. Proposed black hole seeding mechanisms in the early Universe include gravitational collapse of Population III stars \citep[e.g.][]{Madau_2001, Volonteri_2003}, runaway merging \citep[e.g.][]{Portegies_2004Natur, Seth_2008, Devecchi_2009}, or direct collapse \citep[e.g.][]{Loeb_1994, Wise_2008, Regan_2009}. Phases of super-Eddington growth or radiatively inefficient accretion can also help alleviate constraints on heavy black hole seeds \citep[e.g.][]{Madau_2014, Inayoshi_2016, Lupi_2016}. This can be achieved through counter-alignment of the angular momentum vectors of the accretion disc and black hole or uncorrelated mass injection \citep[e.g.][]{King_2006, Zubovas_2021}.

Observations of high-redshift SMBHs and reliable black hole mass measurements are necessary to study quasar demographics and cosmic mass assembly. Large sky surveys have improved sample statistics across a wide range of luminosities and redshifts \citep[e.g.][]{Shen_2011, Milliquas_2015, Rakshit_2020, Wu_2022}, while focused surveys of the ultraluminous quasar subset have directly probed the most massive black hole population and provided stringent constraints on their evolution \citep[e.g.][]{Trakhtenbrot_2011, Lopez_2016_XQ100, Bischetti_2017_WISSH, Schindler_2017_ELQS, Onken_2022_QLF, Cristiani_2023_QUBRICS, Dodorico_2023_XQR30}. 

The most direct quasar demographic studies are based on the quasar luminosity function which has been used to measure active galactic nuclei (AGN) populations and their evolution across a wide range of redshifts \citep[e.g.][]{Willott_2010_QLF, Jiang_2016, Yang_2016, Akiyama_2018_A18, Matsuoka_2018_M18, McGreer_2018, Niida_2020_N20, Onken_2022_QLF, Matsuoka_2023_M23}. Demographic analyses on high-redshift quasars become increasingly difficult due to declining quasar spatial densities, requiring deeper wide-area surveys \citep[e.g.][]{Matsuoka_2016, Matsuoka_2023_M23, Schindler_2023}. Naturally, the faint-end of the luminosity function suffers from incompleteness and the bright-end is affected by small number statistics ($\lesssim$ 1 Gpc$^{-3}$). Another key limitation is the relatively unknown fraction of obscured quasars which can be reddened beyond the typical optical and ultraviolet quasar selection criteria \citep{Ni_2020}, requiring multiwavelength observations for candidate selection and spectroscopic identification.

However, the luminosity function does not directly reflect the mass assembly history, which is more clearly measured by the black hole mass. Follow-up spectroscopic observations are used to estimate black hole masses from quasars identified in large sky surveys using ``single-epoch virial mass estimators'' \citep[e.g.][]{Vestergaard_2006, Shen_2013}, which are based on empirical relationships from reverberation mapping experiments \citep[e.g.][]{Kaspi_2000, Kaspi_2005}. Black hole mass estimates of luminosity-selected samples have enabled measurements of the black hole mass function and its cosmic evolution \citep[e.g.][]{Greene_2005, Vestergaard_2009, Shankar_2009, Willott_2010, Kelly_2012, Kelly_2013, Schulze_2015, Ananna_2022, He_2024_H24}. Translating the flux-limited selection function to the black hole mass function requires assumptions of the Eddington ratio distribution weighted by the quasar duty cycle. Forward-modelling approaches have been developed to better account for uncertainties and biases in measured quantities \citep[e.g.][]{Kelly_2009_BHMF, Schulze_2010, Schulze_2015, Wu_2022_Demo}. One can also constrain a model-independent active black hole mass function directly from observables using volume-weighted binned quasar abundances \citep[i.e. $1/V_{\rm{max}}$ approach;][]{Schmidt_1968}, but the result does not necessarily reflect the intrinsic underlying black hole population.

The work by \citet{He_2024_H24}, hereafter \citetalias{He_2024_H24}, is a study of the $z\sim4$ black hole mass function based on a sample of 52 quasars from the Hyper Suprime-Cam Subaru Strategic Program and 1462 quasars from the Sloan Digital Sky Survey data release 7 quasar catalog \citep{Shen_2011}. By identifying quasar candidates as faint as $i\rm{-mag} = 23.2$ using $g\rm{-band}$ drop-out colours,  \citetalias{He_2024_H24} presents constraints on the black hole mass function down to $\log{M_{\rm{BH}}/M_{\odot}} \sim 7.5$. We use the low-mass abundance constraints of \citetalias{He_2024_H24} in the construction of the $z\sim5$ mass function.

In this work, we study the evolution of the active black hole mass function from $z\sim5$ to $z\sim4$ and discuss how the measured growth rate extrapolates to higher redshifts. Using a spectroscopic follow-up of the most luminous and complete sample of quasars between $4.5 < z < 5.3$ \citep[XQz5;][]{Lai_XQz5} supplemented by literature quasars in the same survey footprint (referred to hereafter as XQz5+), we present measurements of the active black hole mass function at $z\sim5$. We measure the mass evolution observed between the $z\sim5$ and $z\sim4$ mass functions, as well as within the XQz5+ sample itself, which spans $\sim240$ Myr of cosmic time. We discuss the implications of our result on the quasar duty cycle, the spin-dependent radiative efficiency, and the size of black hole seeds at earlier cosmic epochs. 

The content of this paper is organised as follows: in Section \ref{sec:method}, we describe our method, beginning with a description of the $z\sim 5$ quasar sample and its parent survey. We then discuss our approach to measuring and modelling the black hole mass distribution function with the completeness correction of the sample. In Section \ref{sec:results_discussion}, we apply our method to measure the black hole mass function for $z\sim5$ and measure its redshift evolution with literature mass functions at different redshifts. We also discuss measuring the mass evolution within the $z\sim5$ sample using a Monte Carlo mock universe model. We present a summary and conclusion in Section \ref{sec:conclusion}. For this study, we adopt a standard flat $\Lambda$CDM cosmology with H$_{0} = 70$ km s$^{-1}$ Mpc$^{-1}$ and $\left(\Omega_{\rm m}, \Omega_{\Lambda}\right) = \left(0.3, 0.7\right)$. 

\section{Method} \label{sec:method}
The sample used in our analysis, XQz5 \citep{Lai_XQz5}, is based on a survey of southern quasars with unprecedented completeness \citep[][hereafter \citetalias{Onken_2022_QLF}]{Onken_2022_QLF}. From this parent survey, we construct a spectral atlas of ultraluminous $z\sim 5$ quasars composed of optical and near-infrared spectroscopic follow-up observations of the brightest 83 quasars in the \citetalias{Onken_2022_QLF} survey between the redshift range $4.5 < z < 5.3$. The spectroscopic follow-up was performed with the following instruments: SOAR/TripleSpec4.1, VLT/X-shooter, and ANU2.3m/WiFeS, and the reduced data have been made publicly available. A full description of the observations and the data reduction was published by \citet{Lai_XQz5}.

\subsection{Sample completeness}

\begin{figure*}
	\includegraphics[width=0.95\textwidth]{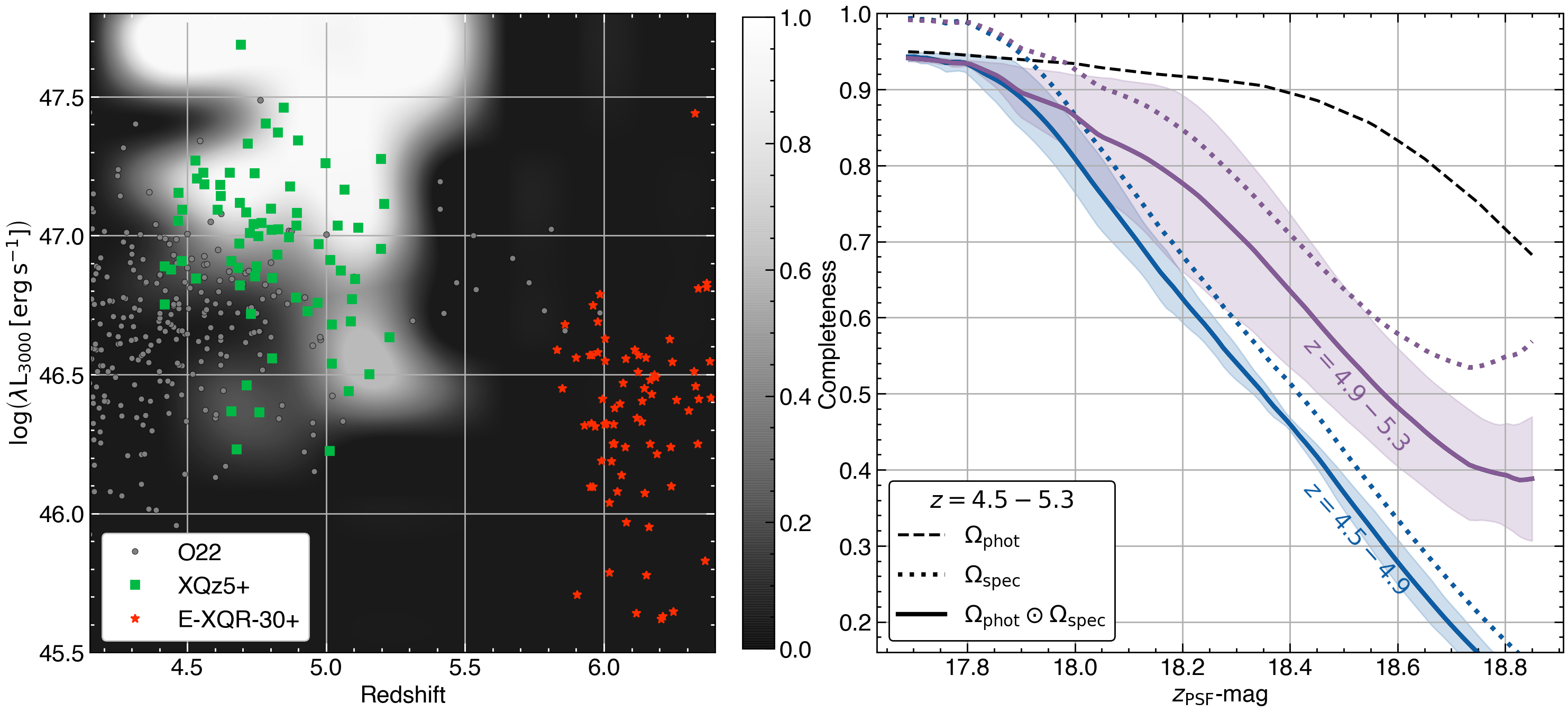}
    \caption[2-D completeness function]{(\textit{Left:}) Completeness function on a 2-D grid of luminosity and redshift, where the photometric and spectroscopic completeness is mapped onto the parameter space in grayscale. The $L_{3000}$ monochromatic luminosity is estimated from the \textit{H}-band magnitude as described in \citetalias{Onken_2022_QLF}. Each quasar in the \citetalias{Onken_2022_QLF} parent sample is shown with a gray point, and the quasars with spectroscopic follow-up that we use in the $z\sim5$ analysis are overplotted as green squares. We estimate the completeness of the spectroscopic follow-up down to $z=5.4$, where the parent sample is assumed to be incomplete. We show E-XQR-30+ with red stars as a high-redshift comparison sample. (\textit{Right:}) Median completeness as a function of $z_{\rm{PSF}}$-mag estimated between the redshift range $4.5 < z < 5.3$, with the spectroscopic completeness binned into $4.5 < z < 4.9$ and $4.9 < z < 5.3$. The total completeness is shown in solid lines with the median absolute deviation across redshifts represented as sheaths. The photometric and spectroscopic completeness, $\Omega_{\rm{phot}}$ and $\Omega_{\rm{spec}}$, are shown with the dashed and dotted lines respectively, where the photometric completeness is defined for the entire redshift range.}
    \label{fig:completeness}
\end{figure*}

The \citetalias{Onken_2022_QLF} parent sample utilises the SkyMapper Southern Survey Data Release 3 \citep[SMSS DR3;][]{Wolf_2018_SkymapperDR1, Onken_2019_SkymapperDR2} in combination with precision astrometry from Gaia DR2/eDR3 as well as infrared photometric suveys from Two Micron All-Sky Survey \citep[2MASS;][]{2MASS}, the VISTA Hemisphere Survey \citep[VHS;][]{VHS} DR6, VISTA Kilo-Degree Infrared Galaxy Survey \citep[VIKING;][]{VIKING} DR5, AllWISE \citep{WISE, ALLWISE}, and CatWISE2020 \citep{Marocco_2021_catwise2020}. The full survey covers 14,486 deg$^2$ of the sky, including all known quasars in the search area as listed in the Milliquas v7.1 update \citep{Milliquas_2015} and 126 newly identified luminous ($M_{1450} < -27$ mag) quasars. The magnitude-dependent completeness, defined in Figure 4 of \citetalias{Onken_2022_QLF} as a function of the SMSS \textit{z}-band, declines from 95\% at $z_{\rm{PSF}}=17.5$ and 90\% at $z_{\rm{PSF}}=18.4$ mag to 78\% at $z_{\rm{PSF}}=18.7$ mag. The survey is assumed to be complete up to $z = 5.4$, where the redshift constraints are imposed by the absence of signal in the Gaia \textit{G}-passband from which astrometric measurements are made \citepalias{Onken_2022_QLF}. 

The full completeness correction in our study, $\Omega(L, z)$, is the Hadamard product of the 1-D magnitude-dependent completeness of \citetalias{Onken_2022_QLF} converted to luminosity, $\Omega_{\rm{phot}}(L,z)$, and the completeness of the spectroscopic follow-up, $\Omega_{\rm{spec}}(L, z)$, described by
\begin{equation}
    \Omega(L, z) = \Omega_{\rm{phot}}(L,z)\odot\Omega_{\rm{spec}}(L, z)\,,
    \label{eq:completeness-corr}
\end{equation}
where $\Omega_{\rm{spec}}(L, z)$ is estimated in a 2-D grid of luminosity and redshift. In the left panel of Figure \ref{fig:completeness}, we show the 2-D completeness function on a grid of $L_{3000}$ monochromatic luminosity and redshift, where $L_{3000}$ is estimated from a composite spectrum of bright $z=1-2$ quasars \citep{Selsing_2016} scaled to the $H$-band magnitude as described in \citetalias{Onken_2022_QLF}. We show the parent sample as grey points and observed sample as green squares. The spectroscopic completeness is determined by the fraction of quasars in \citetalias{Onken_2022_QLF} that are included in our sample with near-IR spectroscopic follow-up, smoothly interpolated over the grid. We also plot a high-redshift comparison sample as red stars, which is a combination of E-XQR-30 \citep{Dodorico_2023_XQR30} augmented by the database of \citet{Fan_2023_ARAA} into a compilation we refer to as E-XQR-30+. 

In the right panel of Figure \ref{fig:completeness}, we show the photometric (dashed line), spectroscopic (dotted lines), and full completeness (solid lines) measured as a function of the $z_{\rm{PSF}}$-magnitude between two bins of redshift over $4.5 < z < 5.3$. The spectroscopic completeness, $\Omega_{\rm{spec}}$, and its uncertainty are measured by the median and median absolute deviation over the redshift range. In the plot, the $z_{\rm{PSF}}$ range is limited to the region where the spectroscopic completeness is well-behaved, at magnitudes of $z_{\rm{PSF}} < 18.85$, approximately corresponding to $\log \lambda L_{3000}/\rm{erg\,s^{-1}} \gtrsim 46.8$ at $z\sim5$. Our sample is more complete at higher redshifts with the median completeness between the two redshift bins deviating by $\sim 0.1$ around $z_{\rm{PSF}} \sim 18.1$.

Our analysis is primarily focused on the high luminosity $(\log \lambda L_{3000}/\rm{erg\,s^{-1}} \gtrsim 46.8)$ quasars between redshifts $4.5 < z < 5.3$, for which we have well-defined completeness. The surveyed sky area contains $55/84$ of the \citetalias{Lai_XQz5} quasars (majority of which are excluded by sky position), as well as 9 quasars from \citet{Trakhtenbrot_2011}, hereafter \citetalias{Trakhtenbrot_2011}, and 8 quasars from \citet{Lopez_2016_XQ100}, hereafter \citetalias{Lopez_2016_XQ100}. The final dataset is composed of 72 quasars. The \citetalias{Trakhtenbrot_2011} and \citetalias{Lopez_2016_XQ100} quasars supplement the lower-luminosity and lower-redshift ends of the sample, respectively. These additional quasars are targets within the footprint of \citetalias{Onken_2022_QLF} which were already observed with spectroscopic follow-up and thus repeat observations were not scheduled for the \citetalias{Lai_XQz5} dataset. We adopt the measured black hole masses from the \citetalias{Trakhtenbrot_2011} and \citet{Lai23_XQ100} analyses, respectively, using the virial estimator and calibration described in Section \ref{sec:bhmf}. We present all 72 quasars used in our demographic analysis in Table \ref{tab:quasar_sample} and we refer to the expanded dataset as XQz5+. The monochromatic $\log \lambda L_{3000}/\rm{erg\,s^{-1}}$ luminosity in Table \ref{tab:quasar_sample} is measured from the near-infrared spectroscopic observations rather than the $H$-band magnitude. The mean difference and standard deviation is $\Delta\log \lambda L_{3000}/\rm{erg\,s^{-1}} = -0.04\pm0.04$. In the following section, we discuss how the completeness correction is used to measure the observed black hole mass distribution function.

\begingroup
\begin{table*}
\caption {\label{tab:quasar_sample} Properties of quasars in XQz5+ used in the demographic analysis.}  
\begin{tabular}{lllccccc}
\hline \hline
ID & RA (J2000) & Dec (J2000) & Redshift & $\log L_{3000}/\rm{erg\, s^{-1}}$ & $\log M_{\rm{BH}}/M_{\odot}$ & $\log L_{\rm{bol}}/L_{\rm{Edd}}$ & Source\\
\hline
000651$-$620803 & 1.71502 & -62.13430 & $4.44$ & $46.787\pm0.029$ & $9.73\pm0.05$ & $-0.45\pm0.05$ & \citetalias{Lopez_2016_XQ100} \\
001225$-$484829 & 3.10424 & -48.80830 & $4.62$ & $47.256\pm0.019$ & $9.80\pm0.05$ & $-0.06\pm0.05$ & \citetalias{Lai_XQz5} \\
001714$-$100055 & 4.31113 & -10.01536 & $5.02$ & $46.547\pm0.023$ & $9.37\pm0.10$ & $-0.34\pm0.11$ & \citetalias{Lai_XQz5} \\
002526$-$014532 & 6.36181 & -1.75905 & $5.06$ & $47.072\pm0.036$ & $9.33\pm0.05$ & $0.23\pm0.06$ & \citetalias{Lai_XQz5} \\
003525$+$004002 & 8.85542 & 0.66750 & $4.76$ & $46.380\pm0.150$ & $8.49\pm0.44$ & $0.38\pm0.46$ & \citetalias{Trakhtenbrot_2011} \\
011546$-$025312 & 18.94274 & -2.88676 & $5.08$ & $46.505\pm0.005$ & $9.51\pm0.07$ & $-0.52\pm0.07$ & \citetalias{Lai_XQz5} \\
013127$-$032059 & 22.86391 & -3.34998 & $5.20$ & $47.108\pm0.029$ & $9.50\pm0.10$ & $0.10\pm0.10$ & \citetalias{Lai_XQz5} \\
013539$-$212628 & 23.91370 & -21.44118 & $4.90$ & $47.385\pm0.011$ & $10.05\pm0.26$ & $-0.18\pm0.26$ & \citetalias{Lai_XQz5} \\
014741$-$030247 & 26.92301 & -3.04659 & $4.80$ & $47.199\pm0.007$ & $10.41\pm0.04$ & $-0.72\pm0.04$ & \citetalias{Lai_XQz5} \\
015618$-$044139 & 29.07906 & -4.69444 & $4.93$ & $46.557\pm0.024$ & $9.28\pm0.11$ & $-0.24\pm0.11$ & \citetalias{Lai_XQz5} \\
020436$-$252315 & 31.15277 & -25.38757 & $4.87$ & $46.965\pm0.016$ & $9.24\pm0.04$ & $0.21\pm0.05$ & \citetalias{Lai_XQz5} \\
021043$-$001818 & 32.67984 & -0.30510 & $4.71$ & $46.040\pm0.080$ & $9.09\pm0.40$ & $-0.56\pm0.41$ & \citetalias{Trakhtenbrot_2011} \\
022112$-$034252 & 35.30259 & -3.71447 & $5.02$ & $46.535\pm0.044$ & $9.05\pm0.17$ & $-0.03\pm0.18$ & \citetalias{Lai_XQz5} \\
022306$-$470902 & 35.77812 & -47.15069 & $5.00$ & $47.297\pm0.018$ & $9.96\pm0.08$ & $-0.17\pm0.09$ & \citetalias{Lai_XQz5} \\
023648$-$114733 & 39.20236 & -11.79268 & $5.20$ & $46.890\pm0.013$ & $9.77\pm0.12$ & $-0.39\pm0.12$ & \citetalias{Lai_XQz5} \\
030722$-$494548 & 46.84538 & -49.76336 & $4.72$ & $47.261\pm0.019$ & $10.07\pm0.07$ & $-0.32\pm0.07$ & \citetalias{Lopez_2016_XQ100} \\
033119$-$074142 & 52.83191 & -7.69525 & $4.73$ & $46.550\pm0.040$ & $8.83\pm0.11$ & $0.21\pm0.12$ & \citetalias{Trakhtenbrot_2011} \\
033703$-$254831 & 54.26273 & -25.80878 & $5.11$ & $46.872\pm0.025$ & $9.64\pm0.10$ & $-0.28\pm0.10$ & \citetalias{Lai_XQz5} \\
040732$-$281031 & 61.88725 & -28.17531 & $4.74$ & $46.675\pm0.010$ & $9.19\pm0.03$ & $-0.03\pm0.03$ & \citetalias{Lai_XQz5} \\
040914$-$275632 & 62.31198 & -27.94248 & $4.48$ & $46.975\pm0.040$ & $9.74\pm0.17$ & $-0.28\pm0.17$ & \citetalias{Lai_XQz5} \\
044432$-$292419 & 71.13547 & -29.40534 & $4.82$ & $46.988\pm0.010$ & $9.89\pm0.06$ & $-0.41\pm0.06$ & \citetalias{Lai_XQz5} \\
045057$-$265541 & 72.73904 & -26.92817 & $4.77$ & $47.021\pm0.027$ & $9.72\pm0.11$ & $-0.21\pm0.11$ & \citetalias{Lai_XQz5} \\
045427$-$050049 & 73.61643 & -5.01375 & $4.83$ & $47.057\pm0.016$ & $9.76\pm0.12$ & $-0.21\pm0.12$ & \citetalias{Lai_XQz5} \\
051508$-$431853 & 78.78722 & -43.31493 & $4.61$ & $47.051\pm0.028$ & $9.69\pm0.34$ & $-0.15\pm0.35$ & \citetalias{Lai_XQz5} \\
052915$-$352603 & 82.31632 & -35.43436 & $4.42$ & $46.690\pm0.018$ & $9.22\pm0.05$ & $-0.04\pm0.05$ & \citetalias{Lopez_2016_XQ100} \\
071431$-$645510 & 108.63081 & -64.91962 & $4.46$ & $47.019\pm0.012$ & $9.61\pm0.07$ & $-0.11\pm0.07$ & \citetalias{Lopez_2016_XQ100} \\
072011$-$675631 & 110.04869 & -67.94215 & $4.62$ & $47.249\pm0.023$ & $10.08\pm0.06$ & $-0.35\pm0.06$ & \citetalias{Lai_XQz5} \\
091655$-$251145 & 139.23200 & -25.19607 & $4.85$ & $47.317\pm0.012$ & $9.77\pm0.05$ & $0.04\pm0.05$ & \citetalias{Lai_XQz5} \\
093032$-$221207 & 142.63577 & -22.20213 & $4.89$ & $47.022\pm0.012$ & $9.90\pm0.06$ & $-0.40\pm0.06$ & \citetalias{Lai_XQz5} \\
095500$-$013006 & 148.75040 & -1.50189 & $4.42$ & $46.814\pm0.021$ & $9.65\pm0.04$ & $-0.35\pm0.05$ & \citetalias{Lopez_2016_XQ100} \\
103623$-$034318 & 159.09895 & -3.72192 & $4.53$ & $46.766\pm0.011$ & $9.46\pm0.03$ & $-0.20\pm0.03$ & \citetalias{Lopez_2016_XQ100} \\
111054$-$301129 & 167.72790 & -30.19164 & $4.78$ & $47.335\pm0.011$ & $10.29\pm0.05$ & $-0.47\pm0.05$ & \citetalias{Lai_XQz5} \\
111520$-$193506 & 168.83470 & -19.58506 & $4.66$ & $47.019\pm0.012$ & $9.71\pm0.07$ & $-0.21\pm0.07$ & \citetalias{Lai_XQz5} \\
113522$-$354838 & 173.84172 & -35.81076 & $4.97$ & $46.952\pm0.008$ & $9.67\pm0.03$ & $-0.23\pm0.04$ & \citetalias{Lai_XQz5} \\
\hline \hline
\end{tabular}
\end{table*}
\endgroup
\setcounter{table}{0}
\begingroup
\begin{table*}
\caption {(Continued)}  
\begin{tabular}{lllccccc}
\hline \hline
ID & RA (J2000) & Dec (J2000) & Redshift & $\log L_{3000}/\rm{erg\, s^{-1}}$ & $\log M_{\rm{BH}}/M_{\odot}$ & $\log L_{\rm{bol}}/L_{\rm{Edd}}$ & Source\\
\hline
120441$-$002149 & 181.17393 & -0.36373 & $5.09$ & $46.585\pm0.023$ & $9.73\pm0.06$ & $-0.66\pm0.07$ & \citetalias{Lai_XQz5} \\
120523$-$074232 & 181.34642 & -7.70907 & $4.69$ & $47.229\pm0.015$ & $10.04\pm0.09$ & $-0.33\pm0.09$ & \citetalias{Lai_XQz5} \\
121402$-$123548 & 183.51130 & -12.59683 & $4.75$ & $46.777\pm0.022$ & $9.66\pm0.09$ & $-0.40\pm0.10$ & \citetalias{Lai_XQz5} \\
121921$-$360933 & 184.83801 & -36.15917 & $4.80$ & $47.046\pm0.014$ & $10.21\pm0.02$ & $-0.67\pm0.03$ & \citetalias{Lai_XQz5} \\
130031$-$282931 & 195.12973 & -28.49195 & $4.71$ & $47.023\pm0.021$ & $9.58\pm0.14$ & $-0.07\pm0.14$ & \citetalias{Lai_XQz5} \\
132853$-$022441 & 202.22366 & -2.41157 & $4.66$ & $46.280\pm0.080$ & $9.08\pm0.24$ & $-0.31\pm0.25$ & \citetalias{Trakhtenbrot_2011} \\
134134$+$014157 & 205.39250 & 1.69939 & $4.69$ & $46.730\pm0.080$ & $9.82\pm0.24$ & $-0.60\pm0.25$ & \citetalias{Trakhtenbrot_2011} \\
140801$-$275820 & 212.00757 & -27.97228 & $4.47$ & $47.170\pm0.057$ & $9.99\pm0.10$ & $-0.34\pm0.11$ & \citetalias{Lai_XQz5} \\
142721$-$050353 & 216.83984 & -5.06477 & $5.09$ & $46.618\pm0.015$ & $9.51\pm0.06$ & $-0.41\pm0.06$ & \citetalias{Lai_XQz5} \\
151443$-$325024 & 228.68260 & -32.84022 & $4.83$ & $47.151\pm0.025$ & $9.36\pm0.04$ & $0.28\pm0.05$ & \citetalias{Lai_XQz5} \\
153241$-$193032 & 233.17252 & -19.50910 & $4.69$ & $46.871\pm0.024$ & $9.35\pm0.05$ & $0.01\pm0.05$ & \citetalias{Lai_XQz5} \\
153359$-$181027 & 233.49907 & -18.17420 & $5.01$ & $46.870\pm0.041$ & $9.75\pm0.13$ & $-0.40\pm0.14$ & \citetalias{Lai_XQz5} \\
155657$-$172107 & 239.23904 & -17.35207 & $4.75$ & $46.901\pm0.009$ & $9.72\pm0.08$ & $-0.33\pm0.08$ & \citetalias{Lai_XQz5} \\
160111$-$182834 & 240.29657 & -18.47638 & $5.05$ & $46.981\pm0.019$ & $9.54\pm0.10$ & $-0.07\pm0.10$ & \citetalias{Lai_XQz5} \\
194124$-$450023 & 295.35245 & -45.00656 & $5.21$ & $47.158\pm0.016$ & $9.86\pm0.07$ & $-0.22\pm0.08$ & \citetalias{Lai_XQz5} \\
205559$-$601147 & 313.99667 & -60.19648 & $4.97$ & $46.961\pm0.015$ & $9.43\pm0.12$ & $0.01\pm0.12$ & \citetalias{Lai_XQz5} \\
205724$-$003018 & 314.35062 & -0.50522 & $4.68$ & $46.830\pm0.080$ & $9.23\pm0.24$ & $0.09\pm0.25$ & \citetalias{Trakhtenbrot_2011} \\
211105$-$015604 & 317.77335 & -1.93449 & $4.89$ & $47.155\pm0.013$ & $9.86\pm0.09$ & $-0.22\pm0.09$ & \citetalias{Lai_XQz5} \\
211920$-$772253 & 319.83676 & -77.38142 & $4.56$ & $46.954\pm0.011$ & $9.96\pm0.14$ & $-0.52\pm0.14$ & \citetalias{Lai_XQz5} \\
215728$-$360215 & 329.36758 & -36.03752 & $4.69$ & $47.672\pm0.001$ & $10.33\pm0.12$ & $-0.17\pm0.12$ & \citetalias{Lai_XQz5} \\
220008$+$001744 & 330.03607 & 0.29580 & $4.80$ & $46.510\pm0.080$ & $8.82\pm0.24$ & $0.18\pm0.25$ & \citetalias{Trakhtenbrot_2011} \\
220158$-$202627 & 330.49408 & -20.44092 & $4.74$ & $47.014\pm0.028$ & $9.65\pm0.09$ & $-0.15\pm0.09$ & \citetalias{Lai_XQz5} \\
221111$-$330245 & 332.79813 & -33.04606 & $4.65$ & $47.231\pm0.011$ & $9.72\pm0.06$ & $-0.00\pm0.06$ & \citetalias{Lai_XQz5} \\
221644$+$001348 & 334.18340 & 0.23001 & $5.01$ & $46.377\pm0.007$ & $9.30\pm0.04$ & $-0.44\pm0.04$ & \citetalias{Lai_XQz5} \\
221651$-$671443 & 334.21664 & -67.24540 & $4.48$ & $46.838\pm0.020$ & $9.79\pm0.07$ & $-0.46\pm0.08$ & \citetalias{Lopez_2016_XQ100} \\
221705$-$001307 & 334.27374 & -0.21870 & $4.68$ & $46.280\pm0.150$ & $8.63\pm0.44$ & $0.14\pm0.46$ & \citetalias{Trakhtenbrot_2011} \\
222152$-$182602 & 335.47037 & -18.43412 & $4.53$ & $47.302\pm0.018$ & $9.85\pm0.10$ & $-0.06\pm0.10$ & \citetalias{Lai_XQz5} \\
222357$-$252634 & 335.99112 & -25.44284 & $4.80$ & $46.942\pm0.026$ & $9.16\pm0.16$ & $0.27\pm0.16$ & \citetalias{Lai_XQz5} \\
222509$-$001406 & 336.28827 & -0.23523 & $4.89$ & $46.700\pm0.040$ & $9.27\pm0.11$ & $-0.08\pm0.12$ & \citetalias{Trakhtenbrot_2011} \\
222612$-$061807 & 336.55173 & -6.30200 & $5.10$ & $46.516\pm0.015$ & $9.35\pm0.05$ & $-0.35\pm0.05$ & \citetalias{Lai_XQz5} \\
222845$-$075755 & 337.18805 & -7.96533 & $5.16$ & $46.323\pm0.032$ & $8.95\pm0.07$ & $-0.14\pm0.08$ & \citetalias{Lai_XQz5} \\
223953$-$055220 & 339.97360 & -5.87223 & $4.56$ & $47.197\pm0.012$ & $9.64\pm0.05$ & $0.04\pm0.05$ & \citetalias{Lopez_2016_XQ100} \\
230349$-$063343 & 345.95496 & -6.56195 & $4.74$ & $47.052\pm0.014$ & $9.89\pm0.09$ & $-0.35\pm0.09$ & \citetalias{Lai_XQz5} \\
230429$-$313426 & 346.12454 & -31.57416 & $4.87$ & $47.181\pm0.013$ & $9.93\pm0.13$ & $-0.26\pm0.13$ & \citetalias{Lai_XQz5} \\
232536$-$055328 & 351.40268 & -5.89114 & $5.23$ & $46.998\pm0.015$ & $9.75\pm0.13$ & $-0.26\pm0.13$ & \citetalias{Lai_XQz5} \\
232952$-$200039 & 352.46988 & -20.01085 & $5.04$ & $47.162\pm0.011$ & $9.59\pm0.06$ & $0.06\pm0.06$ & \citetalias{Lai_XQz5} \\
233435$-$365708 & 353.64703 & -36.95247 & $4.72$ & $46.990\pm0.019$ & $9.51\pm0.29$ & $-0.04\pm0.29$ & \citetalias{Lai_XQz5} \\
233505$-$590103 & 353.77440 & -59.01755 & $4.53$ & $47.262\pm0.010$ & $9.84\pm0.06$ & $-0.09\pm0.06$ & \citetalias{Lai_XQz5} \\
\hline \hline
\end{tabular}
\end{table*}
\endgroup

\subsection{Black hole mass function} \label{sec:bhmf}

Constraining the shape and evolution of the black hole mass function is observationally expensive, due to the requirement of collecting samples with high completeness and obtaining high signal-to-noise spectroscopic data, which are necessary for reliable black hole mass measurements. Furthermore, the completeness correction of the black hole mass function, particularly for lower masses, is not as easily ascertained, as the mass on its own is not sufficient to adequately constrain the intrinsic luminosity, and hence the observed source brightness (on which the completeness estimate relies). Nevertheless, the cosmic evolution of the black hole mass function is a key component in the developing picture of supermassive black hole origins and growth mechanisms. Here, we describe the method used in this study to measure and parameterise the observed black hole mass function.

In the XQz5+ sample, black hole masses are estimated from near-infrared spectroscopic observations of the \mgiil\ broad emission line with the single-epoch virial mass estimator of \citet{Shen_2011}, which is calibrated to the \hbeta\ line in a high-luminosity subset of the local AGN reverberation mapping sample. The single-epoch virial mass equation takes the form,
\begin{equation}
   \left(\frac{M_{\rm{BH,vir}}}{M_{\odot}}\right) = 10^{a} \left[\frac{L_{3000}}{10^{44} \,\rm{erg\, s^{-1}}}\right]^{b} \left[\frac{\rm{FWHM_{\mgii}}}{1000 \,\rm{km\, s^{-1}}}\right]^{c} \,,
   \label{eq:mgii_virial}
\end{equation}
where $L_{3000}$ is the monochromatic luminosity ($\lambda L_{\lambda}$) of the quasar continuum model at 3000\AA\ and FWHM$_{\rm{\mgii}}$ is the measured full-width at half-maximum of the \mgii\ broad line profile. The calibration from \citet{Shen_2011} is $(a,b,c) = (6.74, 0.62, 2.00)$. The intrinsic scatter of the \hbeta\ virial mass estimator is $\sim0.3$ dex compared to their reverberation mapping counterparts \citep{DallaBonta_2020}, while reverberation mapping mass estimates are dispersed around the $M_{\rm BH}-\sigma_{*}$ relation \citep{Bennert_2021} with a $\sim0.4$ dex scatter. Additionally, \mgii-based mass estimates are calibrated to \hbeta, with a dispersion of $0.1-0.2$ dex \citep[e.g.][]{Shen_2012}. As such, the expected systematic uncertainty from single-epoch virial mass estimators is $\sim0.5$ dex, which is often more significant than the measurement uncertainties propagated from the FWHM and continuum luminosity. A more detailed discussion of the systematic uncertainty inherent in single-epoch virial mass measurements is presented in \citet{Lai23_XQ100}, a study of 100 luminous quasars between $3.5 < z < 4.5$.

The binned black hole mass function between $z_{\rm{min}} < z < z_{\rm{max}}$ can be estimated using the $1/V_{\rm{max}}$ method \citep{Schmidt_1968, Avni_1980, Page_2000},
\begin{equation}
    \Phi(M_{\rm{BH}}, z) = \frac{1}{\Delta\log M_{\rm{BH}}} \sum_{i=1}^{N_{\rm{BH}}} \frac{f_{\rm{obs}}^{-1}}{\int_{z_{\rm{min}}}^{z_{\rm{max}}}\Omega(L_i, z) \frac{dV}{dz} dz} \,,
    \label{eq:binned-bhmf}
\end{equation}
where $f_{\rm{obs}} \sim 0.351$ is the sky area coverage of our parent sample, $\Omega(L, z)$ is the 2-D completeness correction from Equation \ref{eq:completeness-corr}, and ${dV}/{dz}$ is the differential comoving volume element. Poisson statistical uncertainties are estimated for each point in the binned mass function. 

The $1/V_{\rm{max}}$ method produces a model-independent and non-parametric estimate of the active black hole mass function. However, we caution that the $1/V_{\rm{max}}$ approach does not consider the error function in the virial black hole mass estimates \citep[e.g.][]{Kelly_2009_BHMF, Shen_2012_Demo, Schulze_2015}. The $1/V_{\rm{max}}$ approach is expected to be biased by a wide error distribution, which artificially broadens the mass function leading to a shallower high-mass slope due to the large systematic errors in the virial $M_{\rm{BH}}$ estimates. The survey selection function is also a function of the quasar luminosity, which does not translate directly to completeness in black hole mass \citep[e.g.][]{Schulze_2010} Therefore, the black hole mass function derived using $1/V_{\rm{max}}$ does not reflect the intrinsic black hole mass function. Numerous other studies have discussed the shortcomings of the $1/V_{\rm{max}}$ approach in more detail and developed more rigorous statistical frameworks \citep[e.g.][]{Kelly_2009_BHMF, Schulze_2010, Schulze_2015, Wu_2022_Demo}. Nevertheless, studies of high-redshift quasar demographics where statistics are sparse continue to use the straightforward and reproducible $1/V_{\rm{max}}$ approach \citep{Matsuoka_2018_M18, Ananna_2022, Matsuoka_2023_M23}, if only as a sanity check to complement other more advanced Bayesian methods \citepalias[e.g.][]{Wu_2022_Demo, He_2024_H24}. In this study, we use the $1/V_{\rm{max}}$ approach to study the differential growth between mass functions separated by a measurable evolution, which would be equally transformed by convolution with homoscedastic error distributions. Therefore, if the error function is homoscedastic and redshift invariant, the measured active black hole mass function \textit{evolution} presented in this study would be independent of the high-mass bias caused by a wide error function.

Double power-laws are often used to parameterise black hole luminosity functions \citep[e.g.][]{Yang_2016, Akiyama_2018_A18, Matsuoka_2018_M18, Niida_2020_N20, Onken_2022_QLF, Matsuoka_2023_M23}, tracing back to earlier studies \citep{Boyle_2000} that observe a flattening of the luminosity function at fainter magnitudes. Motivated by fits to quasar luminosity functions, double power-laws have also been adopted to model black hole mass functions \citepalias[e.g.][]{Li_2023, He_2024_H24} Alternatively, Schechter-like functional forms \citep{Aller_2002}, which resemble galactic stellar mass distributions \citep[e.g.][and references therein]{Baldry_2012, Davidzon_2017}, are also used \citepalias[e.g.][]{Schulze_2015, Ananna_2022, He_2024_H24}. If black hole properties trace host galaxy properties, it is expected that the underlying intrinsic distribution would resemble the Schechter function \citep{Caplar_2015}. In comparison to the double power-law, we find that the Schechter model applied to our data enhances low-mass abundances with a faster exponential turnoff at high masses. In this study, we expect the double power-law to be a better fit to the $1/V_{\rm{max}}$ binned active mass function due to the enhanced spatial abundances at higher masses \citep[see also][]{Schulze_2010} Therefore, we choose to parameterise the observed mass function at redshift $z$ as,
\begin{equation}
    \Phi_{z}(M_{\rm{BH, z}}) = \frac{\Phi(M_{\rm{BH}}^{*})}{(M_{\rm{BH,z}}/M_{\rm{BH}}^{*})^{-(\alpha+1)}+(M_{\rm{BH,z}}/M_{\rm{BH}}^{*})^{-(\beta+1)}} \,,
    \label{eq:dbl_plaw}
\end{equation}
where $M_{\rm{BH}}^{*}$ is the turnover black hole mass, ($\alpha$, $\beta$) are the low- and high-mass slopes, and $\Phi(M_{\rm{BH}}^{*})$ is the normalisation. We translate the observed mass function to other redshifts through the continuity equation described in Section \ref{sec:continuity}. We derive model parameters ($M_{\rm{BH}}^{*}$, $\alpha$, $\beta$) from fitting the $1/V_{\rm{max}}$ binned mass function.

\subsubsection{Eddington ratio distribution}

The Eddington ratio is the luminosity of the quasar as a fraction of its Eddington luminosity, defined as $\lambda \equiv L_{\rm{bol}}/L_{\rm{Edd}}$, where $L_{\rm{Edd}} =  1.26 \times 10^{38} \left({M_{\rm{BH}}}/{M_{\odot}}\right) \rm{erg\, s^{-1}}$. In this study, we estimate $L_{\rm{bol}} = 0.75 \times k_{\rm{BC}} \times L_{3000}$ based on a quasar mean spectral energy distribution (SED) \citep[$k_{\rm{BC}}=5.15$;][]{Richards_2006} and a 25\% anisotropy correction \citep{Runnoe_2012}. We caution that $L_{3000}$-dependent bolometric correction factors as low as $k_{\rm{BC}} \sim 2.5$ for $\log(L_{3000}/{\rm{erg\,s^{-1}}}) = 47.0$ have been suggested by both empirical studies \citep{Trakhtenbrot_2012} and thermal accretion disc models \citep{Netzer_2019}, which primarily cover the ``big blue bump'' of the quasar SED \citep{Shields_1978}. Bolometric corrections to the hard X-ray component are also found to be correlated with the Eddington ratio with differences of an order of magnitude \citep{Vasudevan_2007, Vasudevan_2009}, while optical bolometric corrections show no strong trends with Eddington ratio \citep{Duras_2020}. Another study which integrates the $1\,\mu\rm{m} - 8\,\rm{keV}$ SED is consistent with \citet{Richards_2006}, suggesting $k_{\rm{BC}} = 4.75$ for $\log(L_{3000}/{\rm{erg\,s^{-1}}}) = 47.0$ \citep{Runnoe_2012}. Significant differences in black hole mass or continuum luminosity between samples can have a second-order effect on the bolometric luminosity and hence the derived Eddington ratio by nature of the luminosity or Eddington ratio dependent bolometric correction.

\begin{figure}
	\includegraphics[width=1.0\columnwidth]{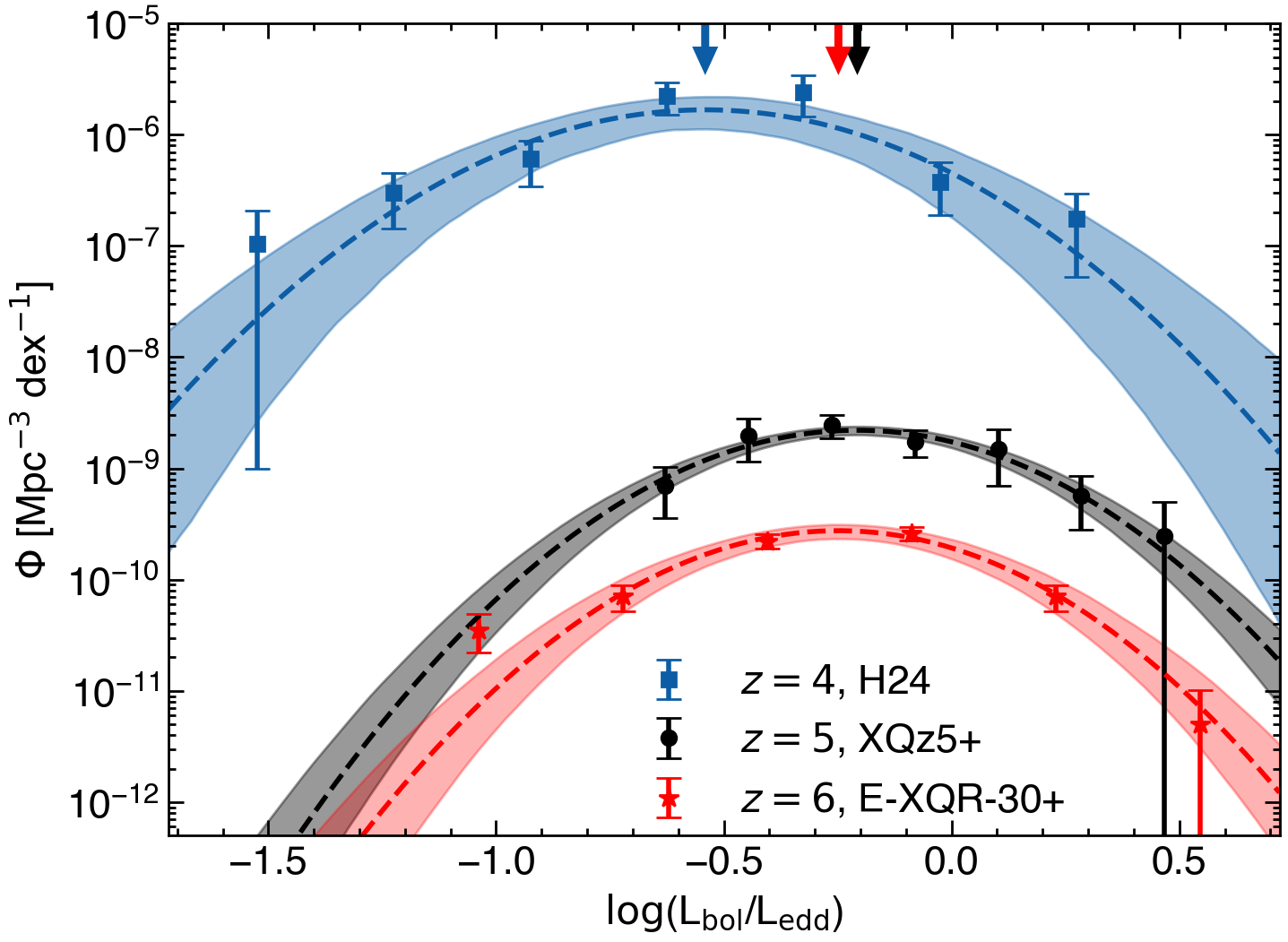}
    \caption[Eddington ratio distribution functions ($4 \gtrsim z \gtrsim 6$)]{Eddington ratio distribution functions of XQz5+ (black circles) alongside \citetalias{He_2024_H24} (blue squares) at $z\sim4$ and E-XQR-30+ (red stars) at $z\sim6$. Mass function model lines share the same colour as the points. The three datasets are derived from surveys which reach different depths. As such, their Eddington ratio distribution functions are not directly comparable. Furthermore, the E-XQR-30+ distribution function is not corrected for completeness. The Eddington ratio distribution of XQz5+ is best fit with a log-normal function, with a mean of $\langle\log\lambda\rangle = -0.21$ and the width of $0.30$ dex. The mean value, indicated with the arrows on the top of the plot, is constrained to $\log(L_{\rm{bol}}/L_{\rm{Edd}}) = -0.54\pm0.06$ for \citetalias{He_2024_H24}, $-0.21\pm0.03$ for XQz5+, and $-0.25\pm0.04$ for E-XQR-30+.}
    \label{fig:ERDF}
\end{figure}

In Figure \ref{fig:ERDF}, we present the Eddington ratio distribution of the sample collated in this study. The binned Eddington ratio abundances are volume-weighted using the $1/V_{\rm{max}}$ method and tabulated in Table \ref{tab:binned_ERDF}. All of the quasars in the sample are accreting with $\lambda > 0.19$. We model the underlying Eddington ratio distribution with a log-normal function,
\begin{equation}
    \rho_{\lambda} = \frac{1}{\sqrt{2\pi}\sigma_{\lambda}}\exp{\left(-\frac{\log{\lambda}-\langle\log\lambda\rangle}{2\sigma_{\lambda}^2}\right)}\,,
\end{equation}
where $\langle\log\lambda\rangle$ is the mean Eddington ratio, and $\sigma_{\lambda}$ is the log standard deviation. Our Eddington ratio distribution, which peaks at $\langle\log\lambda\rangle = -0.21 \pm 0.03$ with $\sigma_{\lambda} = 0.30 \pm 0.02$ dex dispersion, is plotted with those of \citetalias{He_2024_H24} at $z\sim4$ and E-XQR-30+ \citep{Mazzucchelli_2023, Dodorico_2023_XQR30, Fan_2023_ARAA} at $z\sim6$, where each distribution is represented with a mean log-normal model and uncertainty reflected by the shaded region. We note that the three datasets are derived from surveys reaching different depths. Therefore, their Eddington ratio distribution functions are not directly comparable. Furthermore, the E-XQR-30+ Eddington ratio distribution is not adjusted for incompleteness, because the dataset was compiled from inhomogeneous samples such that its selection function is not well-defined. The effect of a flux-dependent selection limit, if applied, is likely to broaden the $z\sim6$ distribution and lower the mean Eddington ratio.

Although the Schechter function is frequently used to model the intrinsic Eddington ratio distribution, a log-normal function is well-matched to the low Eddington turnover, \citep[e.g.][]{Kollmeier_2006, Willott_2010, Shen_2012_Demo, Shen_2019_GNIRS_QSO, Farina_2022}. We note that flux-limited detection limits can emulate the turnover and cause an underestimation of the low-Eddington population in the intrinsic distribution \citep[e.g.][]{Kollmeier_2006, Schulze_2015, Li_2023}. In the absence of low-Eddington constraints, we use the log-normal function to estimate the observed Eddington ratio distribution in our sample. However, as the selection function of the survey is based on the quasar luminosity, which is a product of the black hole mass and Eddington ratio, the Eddington ratio distribution can play a critical role in the mass function completeness \citep[e.g.][]{Kelly_2009_BHMF, Schulze_2010}. Any survey with a fixed flux limit would only include low-mass black holes with the highest Eddington ratios, and exclude high-mass black holes with exceptionally low Eddington ratios. The latter outcome, where supermassive black holes drop out of the parent survey's flux limit due to low accretion rates and/or radiative efficiencies, can affect this study of the differential mass evolution if the intrinsic Eddington ratio distribution evolves over time.

A key question is whether a significant population of quiescent supermassive black holes exists at $z>5$. High-redshift cosmological hydrodynamical simulations \citep[e.g.][]{Li_2007, Sijacki_2009, Bhowmick_2022} find that accretion-driven growth becomes highly efficient by $z\sim6$ and the most massive black holes are accreting close to the Eddington limit. This is in contrast to lower-redshift ($z < 4$) observations \citep[e.g.][]{Barger_2005, Vestergaard_2009, Kelly_2010, Willott_2010} and simulations \citep[e.g.][]{DiMatteo_2008, Sijacki_2015, Volonteri_2016}, which find clear signatures of ``cosmic downsizing'' where the brightest AGN shift to lower black hole masses and the overall energy density production of AGN drops rapidly with decreasing redshift. One possibility to explain this occurrence is the preferential mass starvation of the most massive black holes at low redshift. These studies suggest that the most supermassive black holes at our redshift of interest ($z\sim5$) are likely accreting quickly with high Eddington ratios, implying that the highest mass ($\log M_{\rm{BH}}/M_{\odot} > 9.5$) bins in the $z\sim5$ active black hole mass function could be highly complete in a luminosity-complete survey.

Recent data from the James Webb Space Telescope (JWST) has uncovered a surprisingly abundant population of faint AGN at $z>4$ \citep[e.g.][]{Harikane_2023, Maiolino_2023, Yang_2023_CEERS, Matthee_2024, Silk_2024}, lying above the local $M_{\rm{BH}} - M_{*}$ relation \citep[e.g.][]{Pacucci_2024, Li_2024}. Most of the black holes with mass estimates are reported to be between $M_{\rm{BH}} \sim 10^6 - 10^8$ \citep[e.g.][]{Akins_2023, Goulding_2023, Harikane_2023, Larson_2023, Schneider_2023, Ubler_2023}. The most massive of these high redshift AGN have important consequences for studies of black hole seeding. Compared to extrapolations of the double power-law quasar luminosity function \citep{Niida_2020_N20}, the recent results have increased the abundance of faint AGN by over an order of magnitude \citep[e.g.][]{Harikane_2023, Maiolino_2023, Matthee_2024}. Our study, which focuses on the highest luminosity and most massive quasar subset, is not affected by the recent JWST revelations on the faint AGN population.

\begingroup
\begin{table}
\centering{
\caption[Binned Eddington ratio distribution of XQz5+]{\label{tab:binned_ERDF} Binned Eddington ratio distribution function for XQz5+ at $z\sim5$ estimated by the $1/V_{\rm{max}}$ approach.}  
\begin{tabular}{ccc}
\hline \hline
 $\log L_{\rm{bol}}/L_{\rm{Edd}}$ & N$_{\rm{QSO}}$  & $\Phi(L_{\rm{bol}}/L_{\rm{Edd}})$\\
 & & $10^{-9}$ Mpc$^{-3}$ dex$^{-1}$\\
 \hline
$-0.63$ & 5 & 0.70 $\pm$ 0.34 \\
$-0.45$ & 12 & 1.98 $\pm$ 0.84 \\
$-0.26$ & 24 & 2.44 $\pm$ 0.58 \\
$-0.08$ & 17 & 1.74 $\pm$ 0.48 \\
$0.10$ & 8 & 1.48 $\pm$ 0.78 \\
$0.29$ & 5 & 0.57 $\pm$ 0.29 \\
$0.47$ & 1 & 0.25 $\pm$ 0.25 \\
\hline \hline
\end{tabular}
}
\end{table}
\endgroup

\subsection{Continuity equation} \label{sec:continuity}

The time evolution of an accretion-dominated black hole mass function can be described by the following continuity equation \citep{Shankar_2013_continuity, Tucci_2017},
\begin{equation}
    \frac{\delta \Phi(M_{\rm{BH}}, t)}{\delta t} = -\frac{\delta\left[\langle \dot{M}_{\rm{BH}}\rangle \Phi(M_{\rm{BH}}, t)\right]}{\delta M_{\rm{BH}}}\,,
\end{equation}
where every black hole grows constantly at the mean accretion rate, $\langle \dot{M}_{\rm{BH}}\rangle$. In practice, this picture is complicated by the duty cycle, which is the fraction of black holes in the active state, and merger events, which redistribute mass. To develop a more complete picture, we relate the Eddington ratio to the mass accretion rate by the radiative efficiency, $\epsilon \in (0, 1)$,
\begin{align}
    \dot{M}_{\rm{BH}} = \frac{(1-\epsilon)\lambda}{\epsilon} \frac{L_{\rm{Edd}}}{c^2}\,,
\end{align}
which simplifies to an exponential mass growth with an e-folding time of $450\lambda^{-1}\epsilon(1-\epsilon)^{-1}$ Myr \citep{Salpeter_1964}. In a more general picture, the mean mass accretion rate can be written with the expectation value for Eddington ratio, which is
\begin{equation}
    \langle \dot{M}_{\rm{BH}} \rangle = \int d\log{\lambda} P(\lambda|M_{\rm{BH}}, z) \lambda U(M_{\rm{BH}}, z) \frac{M_{\rm{BH}}}{450\, \rm{Myr}}\frac{(1-\epsilon)}{\epsilon}\,,
\end{equation}
where $P(\lambda|M_{\rm{BH}}, z)$ is the normalised Eddington ratio probability distribution and $U(M_{\rm{BH}}, z) \in [0, 1]$ is the duty cycle. In this study, our approach is to observationally estimate the effective e-folding time under the assumption of time invariance of the radiative efficiency and Eddington ratio distribution as well as a constant duty cycle $U$ over the observed period. We define a dimensionless growth rate scale factor, 
\begin{equation}
    k_{\rm{ef}} \equiv \frac{450\, \rm{Myr}}{t_{\rm{ef}}} = \frac{\langle\lambda\rangle U(1-\epsilon)}{\epsilon}\,,
    \label{eq:kef}
\end{equation}
which behaves as a joint constraint on the unknown parameters ($\epsilon$, $U$), for an underlying Eddington ratio distribution, $P(\lambda|M_{\rm{BH}}, z)$.  

Standard thin accretion disc models predict that the radiative efficiency is a non-linear, but monotonic function of black hole spin. In the local universe, the fiducial value for the radiative efficiency based on the So\l{}tan argument \citep{Soltan_1982_argument} applied to the local black hole mass density is $\epsilon = 0.1$ \citep{Yu_2002}, which corresponds to a black hole spin of $a \sim 0.7$. For individual quasars, observational constraints on the radiative efficiency can be obtained with X-ray reflection measurements or thermal continuum modelling \citep[see review by][]{Reynolds_2019}. Estimates obtained from inhomogeneous quasar samples at higher redshifts are not significantly different from the local fiducial value \citep[e.g.][and references therein]{Capellupo_2015, Vasudevan_2016, Trakhtenbrot_2017, Reynolds_2021}, which may be expected as massive black holes are spun up by prolonged periods of coherent accretion \citep{Dotti_2013, Volonteri_2013, Dubois_2014, Trakhtenbrot_2014, Capellupo_2015, Ananna_2020}. However, radiatively inefficient accretion from uncorrelated flows \citep{King_2005, King_2008} may be a mechanism by which the most extreme mass black holes could be produced at high redshift \citep{King_2006, Zubovas_2021}.

Observational constraints on quasar lifetimes and duty cycle based on clustering suggest that luminous high-redshift (up to $z\sim4-5$) quasars are biased towards rare massive haloes \citep{Shen_2007_clustering, White_2008}. Subsequent modeling efforts predict that the duty cycle must therefore be high, with $U\sim0.5$ at $z\sim4.5$ to $U\sim0.9$ at $z\sim6$ \citep{Shankar_2010}, which is also supported by clustering simulations \citep{DeGraf_2017}. A high duty cycle implies that most of the massive black hole population in the mass function is reflected in observations. However, these results are at odds with measurements from the sizes of quasar proximity zones \citep[e.g.][]{Fan_2006, Eilers_2017, Eilers_2018, Khrykin_2019, Eilers_2020, Morey_2021, Khrykin_2021, Satyavolu_2023}, which often measure lifetimes shorter than $10^{7}$ yrs and sometimes $< 10^{5}$ yrs even at $z\sim6$. One possible explanation for the discrepancy is that quasar light curves are described by a flickering model \citep{Ciotti_2001, Novak_2011, Oppenheimer_2013, Schawinski_2015, Davies_2020, Satyavolu_2023_obscured} instead of a simple light bulb model. Proximity zone studies probe timescales of the most recent emission episode, which can be orders of magnitude shorter than the overall active fraction in a flickering model. Another possibility is that quasars spend a considerable amount of time growing in an obscured phase in their evolution \citep[e.g.][]{Hopkins_2005}, which implies a significant fraction of obscured quasars at high redshift \citep{Davies_2019, Ni_2020, Satyavolu_2023_obscured}, at a much higher ratio than observations at lower redshifts \citep[e.g.][]{Lawrence_2010, Assef_2015}. Thus, the black hole mass function derived in this study would only reflect the active and unobscured subset of the underlying black hole population. 

While black hole mergers do not alter the total mass in the population, they can still modify the shape of the mass function through redistribution. This can potentially affect the inferred evolution from the continuity equation if mergers play a critical role in growing the highest mass quasars over the probed redshift range. Theoretical and simulation-based works generally suggest that black hole mergers may play an important role at early times $(z \gtrsim 9)$ to provide an early boost to black hole masses \citep{Valiante_2016, Bhowmick_2022} and at late times $(z \lesssim 2)$ when galaxies are comparatively gas-starved \citep{Shankar_2013_continuity, Dubois_2014, Kulier_2015}. However, the majority of the assembled black hole mass can be attributed to secular processes \citep{Martin_2018}. We also note that there are numerous studies in this space with a diversity of assembly histories capable of producing the high black hole masses observed at $z\sim6$ \citep{Li_2007, Sijacki_2009, Costa_2014, Feng_2014, Smidt_2018, Valentini_2021, Zhu_2022_first_quasars}, where the differences originate from the adopted accretion and feedback models. This highlights how continued work in this area will require improved observational constraints on high redshift supermassive black holes. 

\section{Results and Discussion} \label{sec:results_discussion}

We apply the model described in Section \ref{sec:method} to the XQz5+ data using the maximum likelihood approach and compare the result to the \citetalias{He_2024_H24} $z\sim4$ mass function. The observed evolution between these samples suggests an effective e-folding time for accretion-dominated growth based on the continuity equation. We compare this value to what we measure internally within our sample through the integrated mass density and we construct a mock universe simulation from the derived model mass functions to estimate Monte Carlo uncertainties. Finally, we discuss the implications of the measured mass evolution on the duty cycle and the radiative efficiency. We also derive the expected mass function at $z\sim6$ and compare with results from the literature. 

\subsection{XQz5+ mass function and evolution}

\begin{figure}
	\includegraphics[width=1.0\columnwidth]{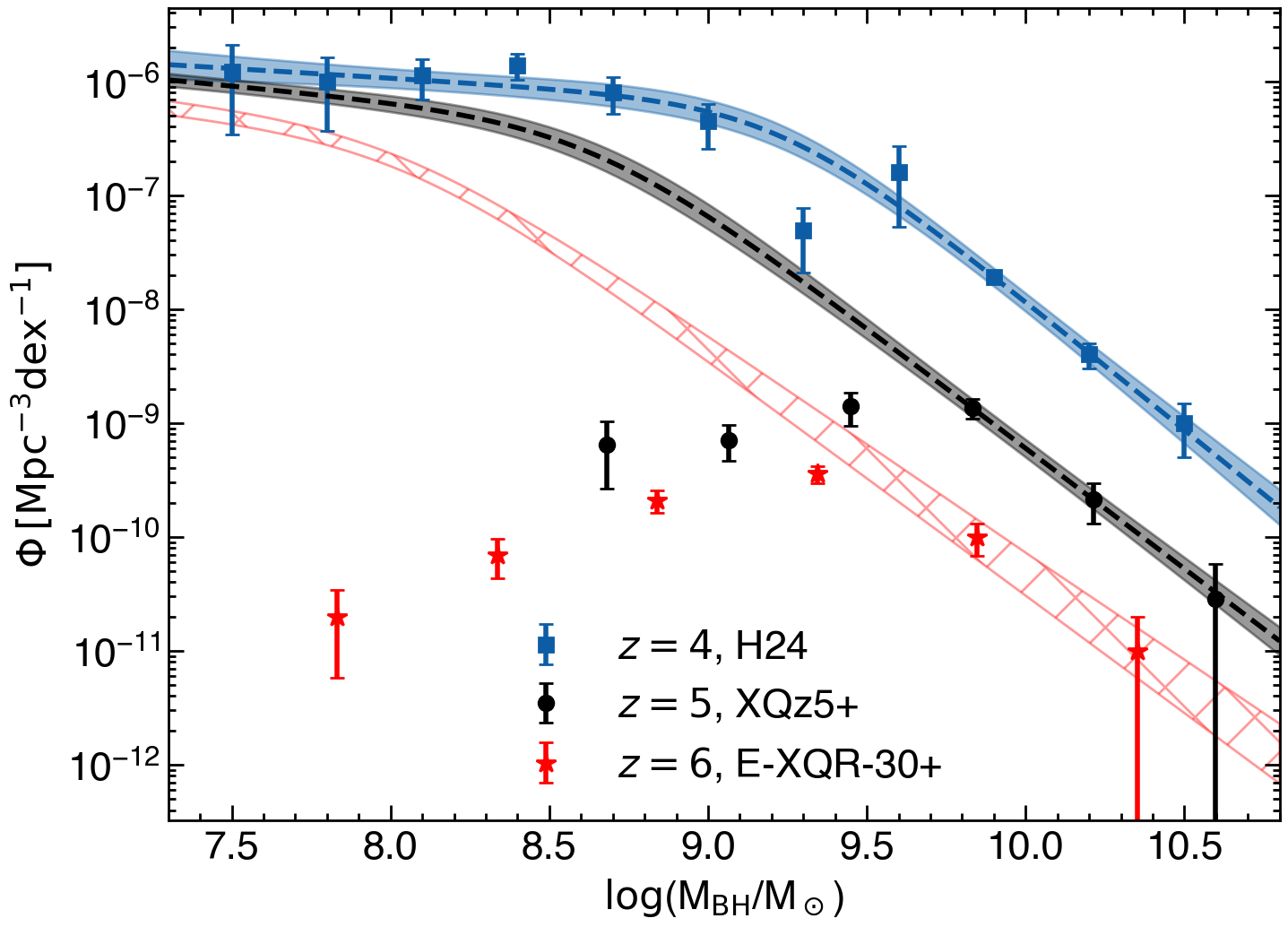}
    \caption[Black hole mass functions for $z=6-4$]{Black hole mass functions of XQz5+ (black circles) with \citetalias{He_2024_H24} (blue squares) at $z\sim4$ and E-XQR-30+ (red stars) at $z\sim6$. The binned values are estimated with the $1/V_{\rm{max}}$ method, but the E-XQR-30+ mass function is not corrected for completeness. We plot the double power-law mass function models and the hatched model is the prediction for $z\sim6$ assuming a constant growth of $k_{\rm{ef}}=1.79$, estimated from $z=5-4$, and extrapolated onto $z\sim6$ from $z\sim5$.}
    \label{fig:BHMF}
\end{figure}

\begingroup
\begin{table}
\centering{
\caption[Binned black hole mass function of XQz5+]{\label{tab:binned_BHMF} Binned black hole mass function for XQz5+ at $z\sim5$ estimated by the $1/V_{\rm{max}}$ approach.}  
\begin{tabular}{ccc}
\hline \hline
 $\log M_{\rm{BH}}/M_{\odot}$ & N$_{\rm{QSO}}$  & $\Phi(M_{\rm{BH}})$\\
 & & $10^{-9}$ Mpc$^{-3}$ dex$^{-1}$\\
 \hline
8.68 & 4 & 0.65 $\pm$ 0.38 \\
9.06 & 9 & 0.71 $\pm$ 0.25 \\
9.45 & 18 & 1.40 $\pm$ 0.45 \\
9.83 & 33 & 1.37 $\pm$ 0.26 \\
10.21 & 7 & 0.21 $\pm$ 0.08 \\
10.60 & 1 & 0.03 $\pm$ 0.03 \\
\hline \hline
\end{tabular}
}
\end{table}
\endgroup

In this section, we first discuss the XQz5+ $z\sim5$ binned black hole mass function. We then discuss two methods by which the dimensionless growth rate, $k_{\rm{ef}}$ in Equation \ref{eq:kef}, is estimated, either from redshifts $z=5-4$ or within $z\sim5$. We compare the results and use a mock universe simulation to estimate the Monte Carlo uncertainties.

In Figure \ref{fig:BHMF}, we present the binned mass function of XQz5+, measured with the $1/V_{\rm{max}}$ method as described in Equation \ref{eq:binned-bhmf}. The black hole mass range of XQz5+ is subdivided into six equally-spaced bins in log-scale, where each bin is occupied by between $N_{\rm{QSO}} = 1-33$ quasars. We present the tabulated abundances in Table \ref{tab:binned_BHMF} and each point is represented by the central black hole mass of the bin. The resulting XQz5+ black hole mass function exhibits an artificial turnover at $\log{(\rm{M_{BH}}/\rm{M_\odot})} \lesssim 9.5$ due to the sample incompleteness. The mass function is resistant to changes in the number of bins. Therefore, we choose six bins to adequately resolve the position of the turnover while simultaneously maximising the number of quasars in each bin. The results of subsequent analyses are insensitive to the binning strategy. 

In order to constrain the low-mass end of the XQz5+ mass function, we refer to \citetalias{He_2024_H24}, which has constrained the mass function at $z\sim4$ in equally spaced bins down to $\log{(\rm{M_{BH}}/\rm{M_\odot})} \sim 7.5$. We first fit their binned black hole mass function for their combined sample using the double power-law in Equation \ref{eq:dbl_plaw} and present the optimised parameters in Table \ref{tab:BHMF_model_params} with associated uncertainties. The $z\sim4$ mass function from \citetalias{He_2024_H24} is then transformed to $z\sim5$ with the continuity equation, optimising the mass accretion rate such that the transformed model fits the observed XQz5+ mass function at $\log{(\rm{M_{BH}}/\rm{M_\odot})} \gtrsim 9.5$. After this, the double power-law is fit to the three most massive XQz5+ bins with the low-mass end constrained by the transformed $z\sim4$ data. The optimised XQz5+ mass function parameters are presented in Table \ref{tab:BHMF_model_params} with associated uncertainties.

\subsubsection{Growth between $z=5-4$} \label{sec:growth5-4}

Both the \citetalias{He_2024_H24} $z\sim4$ and XQz5+ $z\sim5$ binned mass function with their respective models are presented in Figure \ref{fig:BHMF}. The difference in the slopes of these mass functions is not statistically significant. However, the turnover mass, $M_{\rm{BH}}^{*}$, has translated horizontally and the normalisation of the mass function also increases with cosmic time. This observable evolution from the $z\sim5$ to $z\sim4$ mass functions implies an effective growth scale factor of $k_{\rm{ef}} = 1.79 \pm 0.01$, which is fit to the highest mass data with the lowest uncertainties in \citetalias{He_2024_H24}. The $k_{\rm{ef}}$ error term does not include uncertainties in the mass functions, their completeness corrections, or cosmic variance, which are the dominant sources of uncertainty. To account for this, we model the systematic error from mock universe simulations in Section \ref{sec:mock-universe}.

\subsubsection{Growth within $z\sim5$}

As a sanity check, we independently measure the mass evolution within the XQz5+ sample. We bin the redshift range from $z=4.5 - 5.2$ into $N_{\rm{bin}}$ shells of equivalent comoving volume. The redshift range, covering $\sim200$ Myr, is selected where the completeness is more reliable and less affected by the details of interpolation. In each bin, we measure the completeness-corrected total mass of the $N_{\rm{BH}}$ most massive black holes. The critical assumption in this method is that the selected black holes in each redshift bin reflect a random sampling from an underlying population of the most extreme supermassive black holes that is growing with a mean population rate that then evolves into subsequent redshift bins. In principle, if the assumption holds true, the result would be more reliable for increasing quantities of $N_{\rm{bin}}$ shells and $N_{\rm{BH}}$ quasars. However, due to the limited size of the quasar sample, we fit the standard exponential accretion growth for all combinations of $N_{\rm{bin}} = [4, 6]$ and $N_{\rm{BH}} = [4, 10]$. The lower limit of $N_{\rm{bin}} = 4$ and $N_{\rm{BH}} = 4$ is least sensitive to completeness corrections, while the upper limit of $N_{\rm{bin}} = 6$ and $N_{\rm{BH}} = 10$ utilises almost all of the quasars in the XQz5+ sample. With this independent method, we find that quasars across the XQz5+ sample are evolving with $k_{\rm{ef}} = 1.89 \pm 0.31$, representing the mean and standard deviation of all fits. The result is consistent with the observed growth in the $\sim360$ Myr from redshifts $z = 5-4$ in Section \ref{sec:growth5-4}. 

\begingroup
\begin{table*}
\caption[Model black hole mass function parameters]{\label{tab:BHMF_model_params} Optimised fit parameters for the black hole mass function double power-law in Equation \ref{eq:dbl_plaw}. We fit the model to datasets spanning different redshift ranges. The bolded parameters represent the $z\sim5$ mass function maximum likelihood fit from this study.}  
\begin{tabular}{cccccc}
\hline \hline
Redshift & Dataset & $\Phi(M_{\rm{BH}}^{*})$ & $M_{\rm{BH}}^{*}$ & $\alpha$ & $\beta$ \\
 & & $10^{-7}$ Mpc$^{-3}$ dex$^{-1}$ & $10^{x}M_{\odot}$ & & \\
\hline
$3.50 < z < 4.25$ & \citet{He_2024_H24} & $6.6\pm2.1$ & $9.23\pm0.09$ & $-1.17\pm0.12$ & $-3.26\pm0.15$ \\
$\mathbf{4.50 < z < 5.30}$ & \textbf{This work} & $\mathbf{4.6\pm1.3}$ & $\mathbf{8.64\pm0.11}$ & $\mathbf{-1.26\pm0.09}$ & $\mathbf{-3.12\pm0.11}$ \\
\phantom{*}$5.80 < z < 6.40$* & \citet{Dodorico_2023_XQR30, Fan_2023_ARAA} & $3.7\pm1.3$ & $8.10\pm0.11$ & $-1.31\pm0.11$ & $-3.04\pm0.09$ \\
\hline \hline
\multicolumn{6}{l}{\footnotesize
*Fit to mock universe model, with uncertainties adopted from the $z\sim5$ model.}
\end{tabular}
\end{table*}
\endgroup

\subsubsection{Mock universe model} \label{sec:mock-universe}
To test the robustness of the $k_{\rm{ef}}$ measurements, we simulate mock universes and estimate the Monte Carlo uncertainty. We anchor the simulator's reference mass function to the $z\sim5$ model. The simulator functions by stepping through thin shells of comoving volume over a defined redshift range. We use the continuity equation to evolve the model mass function to the mean shell redshift for a prescribed $k_{\rm{ef}}$, and an integrated number of quasars above a chosen mass threshold are randomly generated within each shell based on the probability distribution. The simulator reliably reproduces the reference $z\sim5$ mass function with the expected Poissonian noise and predicts its evolution across redshifts. 

We use the mock universe simulator to estimate the Monte Carlo uncertainty from measuring the mass evolution within the XQz5+ sample which covers $\sim200$ Myr of time, finding a systematic uncertainty of $\sigma_{k_{\rm{ef}}} = 0.5-0.8$ due to Poisson variance from small number statistics. The value of $\sigma_{k_{\rm{ef}}}$ is sensitive to the width of the redshift range under consideration, where performing a measurement across a $\sim700$ Myr width results in $\sigma_{k_{\rm{ef}}} \sim 0.2$. The uncertainty is also dependent on the number of quasars counted in each bin, but does not change significantly with the number of bins, implying $\sigma_{k_{\rm{ef}}} \propto (\sqrt{N_{\rm{BH}}} \Delta T)^{-1}$.

Furthermore, we determine the expected Monte Carlo uncertainty from measuring the $k_{\rm{ef}}$ derived from $z\sim5$ to $z\sim4$, finding a median absolute deviation of $\sigma_{k_{\rm{ef}}} \sim 0.06$. Deriving the mass evolution factor between the XQz5+ and \citetalias{He_2024_H24} dataset, each covering $\sim240$ and $\sim360$ Myr respectively with $\sim90$ Myr separation, is less sensitive to cosmic variance than using the XQz5+ sample on its own. 

The experiments performed with the mock universe simulator produce estimates of the uncertainty in the black hole mass growth measured from $z=5-4$ and within $z\sim5$ due to cosmic variance. Whichever the method, the uncertainty derived from the mock universe experiments is dominant over the measurement error and we find that this additional systematic uncertainty is most sensitive to the redshift coverage of the sample. We use the results of these experiments to define the statistical significance of our $k_{\rm{ef}}$ measurements.

\subsection{Effective cosmic mass growth} \label{sec:effective_growth}
The dimensionless growth factor, defined in Equation \ref{eq:kef}, is measured to be $k_{\rm{ef}} = 1.79 \pm 0.01$ from the evolution between the $z\sim5$ (XQz5+) and $z\sim4$ (\citetalias{He_2024_H24}) mass functions or $k_{\rm{ef}} = 1.89 \pm 0.31$ from the mass evolution within the XQz5+ sample using the integrated mass of the $N_{\rm{BH}}$ most massive in a range of redshift bins. The expected Monte Carlo uncertainty derived from the mock universe experiment is $\sigma_{k_{\rm{ef}}} > 0.5$ for the mass evolution within XQz5+ and $\sigma_{k_{\rm{ef}}} \sim 0.06$ for the change in the mass function from the $z\sim5$ to $z\sim4$ quasar samples. There is no evidence of a change in black hole mass growth rate within the $z\sim5$ sample and from $z=5-4$, due to the high uncertainty in the $k_{\rm{ef}}$ mass evolution within XQz5+. Therefore, our best estimate for the mass growth of supermassive black holes about $z\sim5$ is $k_{\rm{ef}} = 1.79\pm0.06$.

\begin{figure}
	\includegraphics[width=1.0\columnwidth]{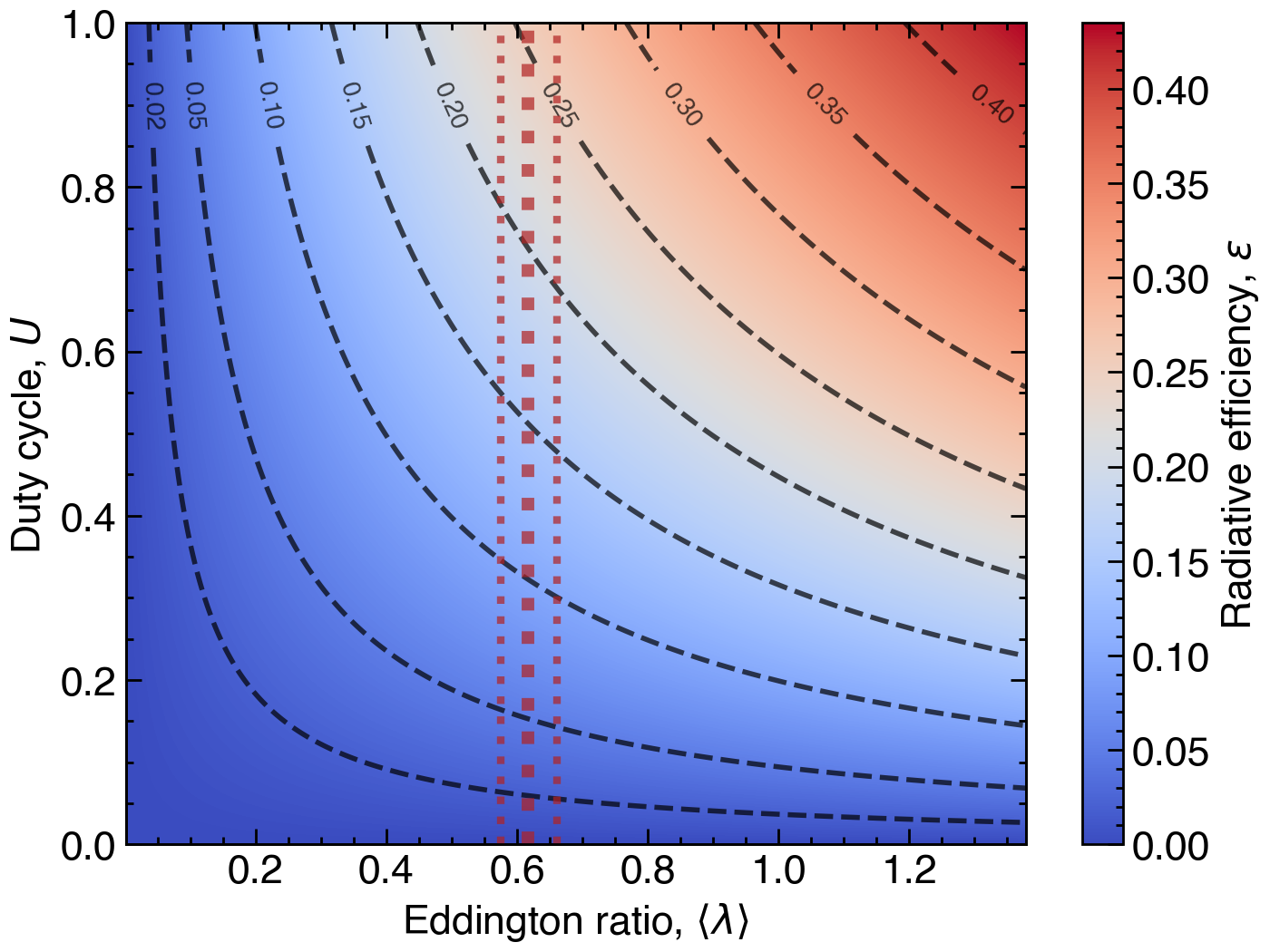}
    \caption[Joint constraints on duty cycle and radiative efficiency]{Surface plot showing the relationship between the Eddington ratio, duty cycle, and radiative efficiency satisfying Equation \ref{eq:kef} for $k_{\rm{ef}} = 1.79$. We overplot dashed radiative efficiency contours and the uncertainty on the expected value of the XQz5+ Eddington ratio log-normal model as the solid line, which is described by a mean and width of $\langle\log\lambda\rangle = -0.21 \pm 0.03$ dex, shown as vertical dotted lines.}
    \label{fig:surface}
\end{figure}

Based on these results, we present a surface plot in Figure \ref{fig:surface}, adopting a $k_{\rm{ef}} = 1.79$ prescription. The plot shows joint constraints on the duty cycle and radiative efficiency for an estimated mean Eddington ratio in the underlying population. Contours of constant radiative efficiency are labelled and plotted as dashed lines. We overplot the expected value, $\langle \lambda \rangle$, of the log-normal Eddington ratio model for XQz5+, marking the mean and standard error with dotted lines. The expectation value is well-constrained to $\langle\log\lambda\rangle = -0.21 \pm 0.03$, or $\langle\lambda\rangle = 0.62\pm0.04$. 

Assuming a fixed $k_{\rm{ef}} = 1.79$ evolution in Figure \ref{fig:surface}, the growth between the $z=5-4$ mass functions suggests that the underlying population of black holes could have mean radiative efficiencies of $\epsilon < 0.32$ ($3\sigma$ of the $\langle\lambda\rangle$ scatter), depending on the duty cycle. However, because $\epsilon$ is a slow function of the dimensionless black hole spin, $a_{*}$, only the highest spin black holes are expected as radiatively efficient as $\epsilon > 0.30$. The results are also consistent with the fiducial $\epsilon = 0.10$ efficiency for a duty cycle of $U\sim0.33$. Alternatively, if the average spin is biased towards low values ($0 \lesssim |a_{*}| \lesssim 0.3)$ expected for chaotic accretion \citep[e.g.][]{King_2008, Berti_2008}, then the duty cycle would be in the range $U = 0.18-0.22$.

If the high-redshift massive black hole population duty cycle is as low as $U\sim0.1$, suggested by the simple light bulb model in proximity zone studies \citep[e.g.][]{Khrykin_2021}, then the observed growth would require the accretion radiation feedback to be inefficient $(\epsilon \sim 0.03)$, which implies that the accretion disc and the black hole spin are counter-aligned. This scenario is plausible if the massive black hole population primarily grows through short episodes of uncorrelated flows \citep{King_2006}, which prevents the black hole from spinning up \citep[e.g.][]{King_2005, Lodato_2006}. Spin misalignment, expected from infalling material that is insensitive to the spin direction of the inner accretion disc, could also result in severely warped structures in the accretion disc \citep[e.g.][]{Chatterjee_2020}, which can obscure the broad-line region \citep{Lawrence_2010}. This is also consistent with the prediction of a significant population of obscured quasars at high redshift \citep[e.g.][]{Davies_2019}, which we discuss further in Section \ref{sec:over-completeness}. Low radiative efficiencies can also be the result of photon trapping in supercritically accreting thick discs \citep[e.g.][]{Ohsuga_2002, Wyithe_2012}.

Conversely at high duty cycles, the measured growth is consistent with a population of quasars that have been spun up by coherent accretion flows over long episodes. Such high duty cycles are consistent with clustering studies \citep[e.g.][]{Shen_2007_clustering, White_2008}, but the constraints on the inferred lifetimes are weak owing to uncertainties in models of quasar-hosting dark matter haloes \citep[e.g.][]{Shen_2009, Cen_2015}. Additional clarity on the high-redshift quasar duty cycle will enable firmer conclusions on the implied black hole spin, based on the observed mass evolution. 

\subsection{Comparison to $z\sim6$} \label{sec:compare_z6}

\begin{figure}
	\includegraphics[width=1.0\columnwidth]{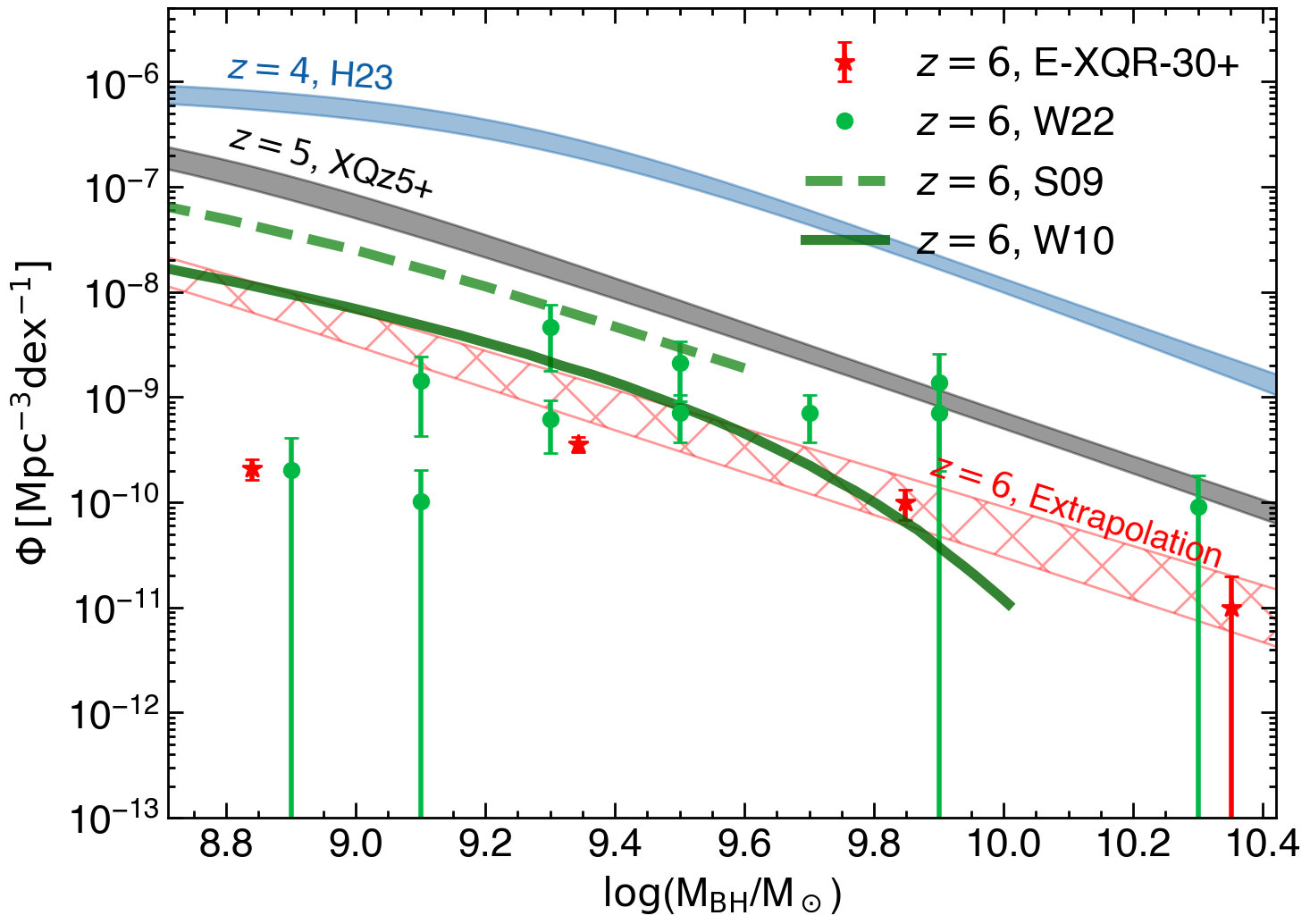}
    \caption[Literature comparison of $z\sim6$ mass functions]{Literature estimates of the $z\sim6$ active black hole mass function compared to the mass function models from Table \ref{tab:BHMF_model_params} at redshifts $z=6-4$, with the same identifying characteristics as in Figure \ref{fig:BHMF}. The active black hole mass functions of \citet{Willott_2010} (green solid line), \citet{Shankar_2009} (green dashed line), \citet{Wu_2022_Demo} (green points), and E-XQR-30+ (red stars) are shown. In some of the \citet{Wu_2022_Demo} mass bins, there are two spatial density estimates, because they consider two quasar samples, each with their own completeness function. We have not corrected E-XQR-30+ for completeness due to heterogeneity in its sample construction.}
    \label{fig:BHMFz6}
\end{figure} 

In Figure \ref{fig:BHMF}, we show the $z\sim6$ mass function measured from the E-XQR-30+ database using the $1/V_{\rm{max}}$ approach without applying completeness corrections. The predicted $z\sim6$ model from our study, represented with the hatched region, is extrapolated from the dimensionless growth factor of $k_{\rm{ef}} = 1.79 \pm 0.06$ derived between $z=5-4$. The optimised $z\sim6$ double power-law fit parameters are presented in Table \ref{tab:BHMF_model_params} with parameter uncertainties adopted from the $z\sim5$ model.

We remark that despite the absence of a completeness correction, the two highest mass binned points of E-XQR-30+ are consistent with the model expectation. Taken at face value, this result would imply that E-XQR-30+ is highly complete $(\Omega(L, z)f_{\rm{obs}} > 0.5)$ at the high mass end and it disfavours scenarios where the black holes evolve more rapidly between $z=6-5$ than $z=5-4$. This observation is at odds with the general expectation that black holes in the early Universe need to evolve quickly to reach their observed masses and are decelerating in their growth as seen in the cosmic downsizing phenomenon \citep[e.g.][]{Barger_2005}. 

We compare the predicted $z\sim6$ black hole mass function model with several other empirical studies of quasar demographics at $z\sim6$ \citep{Shankar_2009, Willott_2010, Wu_2022_Demo}. Compared to semi-analytical or simulation-based models \citep[e.g.][]{Li_2007, Kulier_2015, Amarantidis_2019, Piana_2021}, the methods used to derive these empirical mass functions are more directly comparable to the approach adopted in this study. In Figure \ref{fig:BHMFz6}, we overplot the three literature $z\sim6$ active black hole mass functions with \citetalias{He_2024_H24} ($z\sim4$), XQz5+ ($z\sim5$), and the extrapolated $z\sim6$ black hole mass functions presented in Table \ref{tab:BHMF_model_params}. We also plot the uncorrected binned mass function derived from the E-XQR-30+ database, showing that the literature mass functions generally predict higher quasar abundances. E-XQR-30+ is a much larger heterogeneously constructed sample of quasars with \mgii-based black hole mass estimates and it serves as a sensible lower-bound for the $z\sim6$ quasar space density. The mass functions from \citet{Shankar_2009} and \citet{Willott_2010} have been corrected to the active black hole mass function by removing their duty cycle and obscuration fraction corrections. At the high mass end, we observe that the \citet{Wu_2022_Demo} $z\sim6$ spatial densities are of similar magnitude as our XQz5+ $z\sim5$ sample while the \citet{Shankar_2009} mass function lies above our $z\sim6$ extrapolated model for all masses. In the following discussion, we explore each of the $z\sim6$ literature mass functions in detail.

The mass function from the \citet{Shankar_2009} study was the result of models fit to the local black hole mass function, the X-ray background, and the luminosity functions at a variety of redshifts up to $z\sim6$. The black hole masses are translated to luminosities by assuming a constant Eddington ratio of $\lambda = 0.4$ and radiative efficiency of $\epsilon = 0.065$, while the duty cycle $U(z=6, \log{M_{\rm{BH}}=9.5})=0.5$ is a function of mass and redshift. They have also adopted a correction for the obscured fraction, which is sensitive to AGN luminosity \citep[e.g.][]{Lawrence_1991, Willott_2000, Ueda_2003, Hasinger_2008, Merloni_2014, Ueda_2014}, and they used the luminosity-dependent observable fraction, parameterised with a power-law by \citet{Hopkins_2007}. We invert the obscuration correction to derive the \citet{Shankar_2009} active black hole mass function. In the \citet{Shankar_2009} reference model, the duty cycle is relatively flat for high mass $(\log(M_{\rm{BH}}/M_{\odot}) > 9.0)$ black holes, falling from $U(6, 9.5)=0.5$ to $U(5, 9.5)=0.2$ and subsequently $U(3, 9.5)=0.07$. Under this model, much of the mass assembly would already have occurred at even higher redshifts in order for the results to match the local black hole mass function. Thus, in light of the strong duty cycle evolution and difference in radiative efficiency, it is not surprising that the spatial densities in the \citet{Shankar_2009} mass function is higher at all masses than our extrapolated mass function. 

The analysis from \citet{Willott_2010} assumes $\epsilon = 0.09$, a distribution of Eddington ratios centred at $\lambda = 0.6$ with a dispersion of $0.3$ dex, and a uniform duty cycle distribution between $U=0.5-1.0$. This model is fit to the \citet{Willott_2010_QLF} $z\sim6$ luminosity function based on 40 quasars. Similar to \citet{Shankar_2009} study, an additional correction factor for obscured quasars was applied, which we have inverted to estimate the active black hole mass function. The \citet{Willott_2010} black hole mass function was determined to be considerably lower than the \citet{Shankar_2009} mass function by a factor of $\sim3$. We find that \citet{Willott_2010} is consistent with the extrapolated $z\sim6$ mass function in this study at lower masses, but the high-mass slope is steeper and underestimates the abundance of massive quasars. 

The \citet{Wu_2022_Demo} mass function utilised a sample of 29 quasars with mostly \mgii-based virial black hole mass estimates and just over 100 quasars with $M_{1450}$ estimates from a fixed power-law continuum extrapolation. The quasars with black hole mass estimates are split into two samples \citep[see][]{Jiang_2016}, hence there can be two estimates for the spatial density in the same mass bin. At high masses, the \citet{Wu_2022_Demo} binned spatial density is more consistent with the $z\sim5$ spatial density than with other $z\sim6$ mass functions, but the two highest mass bins which appear to deviate from the mass function of \citet{Willott_2010}, contain only 5 quasars, including J0100$+$2802 $(z=6.3)$, which is the sole occupant of the highest mass bin. The derived spatial densities are then boosted by large correction factors. The black hole mass of J0100$+$2802 adopted in \citet{Wu_2022_Demo} is arguably mildly overestimated based on new James Webb Space Telescope spectra \citep{Eilers_2023_J0100}. Furthermore, no other object in the E-XQR-30+ database, which includes quasars over a much larger sky area, has a measured luminosity comparable to J0100$+$2802, which is more luminous by a factor of ten over the second most luminous quasar. This makes J0100$+$2802 a one-of-a-kind outlier at its redshift. 

Furthermore, while the redshift coverage of \citet{Wu_2022_Demo} extends to $z=6.42$, the space between $6.31 < z < 6.42$ is void of any quasars and two of the three quasars at $z>6.3$ are among the four most massive quasars in the \citet{Wu_2022_Demo} sample such that excluding quasars at the under-complete $z>6.3$ space has a dramatic effect on the mass function in the highest mass bins. By excluding quasars at $z>6.3$, the \citet{Wu_2022_Demo} binned mass function at $\log(M_{\rm{BH}}/M_{\odot}) > 9.2$ becomes consistent within $1\sigma$ with the $z\sim6$ extrapolated model.

A mass growth factor of $k_{\rm{ef}} = 1.5-1.8$ can be measured between the \citet{Wu_2022_Demo} $z\sim6$ and the \citetalias{He_2024_H24} $z\sim4$ binned mass functions. Although the result is consistent with the $z=5-4$ growth rate, the \citet{Wu_2022_Demo} completeness corrections are highly uncertain (L. Jiang, priv. comm.). 

In the following sections, we list and discuss several limitations in our study that can affect the measured growth rate, the extrapolation to $z\sim6$, and explanations for discrepancies between mass functions from different studies.

\subsubsection{Overestimated completeness?} \label{sec:over-completeness}
It is possible that the high flux completeness, $\Omega(L, z)$, of our parent sample at $z\sim5$ does not translate directly to a high completeness in black hole mass as previously assumed. This would be the case if there exists a significant fraction of obscured, quiescent, or low-Eddington quasars with massive black holes. If the active fraction also evolves over the redshift range of $z=6-4$, then the differential correction across mass functions at different redshifts would also capture this time-dependence, limiting the ability to interpret the observed growth in terms of accretion properties. There are indications that the obscured fraction is more significant at higher redshifts than at lower redshifts \citep[e.g.][]{Davies_2019, Vijarnwannaluk_2022}, which would cause the mass functions to diverge. Clustering studies suggest that changes to the obscuration fraction may be secondary to the redshift evolution in the duty cycle \citep{Porciani_2004, Porciani_2006, Shen_2007_clustering}, which describes a general decline in quasar activity over the age of the universe across all masses. The high space density in the massive black hole population observed at $z\sim4$ \citep{He_2024_H24} and the consistency in the mass function slope in the $z=6-4$ redshift range suggest that the most massive black holes have yet to turn off and move to lower luminosities, as described by cosmic downsizing \citep[e.g.][]{Barger_2005, Vestergaard_2009, Kelly_2010}. In order to reliably correct for these effects, more work is needed to understand the population of obscured quasars and the quasar duty cycle at high redshift.

\subsubsection{Overestimated black hole masses?} \label{sec:overBHmass}
Overestimated black hole masses can have a significant effect on the black hole mass function. Black hole masses can be overestimated in a systematic fashion when measurements of the most massive black holes are the result of the long tail of the virial black hole mass estimate error function or when mass estimates are derived from different emission-lines with separate calibrations. We first remark that the \citetalias{He_2024_H24} $z\sim4$ black hole masses are derived from the broad \civ\ emission-line, whereas the $z\sim (5,6)$ samples use the \mgii\ line. The \civ\ line is more likely to be affected by non-virial motions \citep[e.g.][]{Proga_2000, Richards_2011, Shen_2012, Trakhtenbrot_2012, MejiaRestrepo_2018, Saturni_2018}, which motivated the addition of correction factors to the mass calibration accounting for, among others, the \civ\ blueshift \citep[e.g.]{Coatman_2016, Coatman_2017} and the peak flux ratio of the \civ\ line with other UV emission lines \citep[e.g.][]{Runnoe_2013, Mejia-Restrepo_2016}. \citetalias{He_2024_H24} also investigated an alternative mass calibration \citep{Park_2017}, which sought to compensate for the peculiarities of \civ. This led to a much tighter black hole mass distribution centred at lower masses, which would imply a steeper bright-end slope and lower-mass turnover in the $z\sim4$ mass function. Our analysis would then have overestimated the growth in the most massive black holes between $z=5-4$. However, \citetalias{He_2024_H24} ultimately disfavoured this alternative scenario on account of the lack of high-mass calibrations sources and the abandonment of a virial-like dependence on the line width.

Black hole masses estimated from single-epoch virial mass estimates depend on a high-luminosity extrapolation of the radius-luminosity relationship from reverberation mapping experiments \citep{Mclure_2004, Shen_2011}. Reverberation mapping campaigns focused on high-redshift and high-luminosity quasars are expensive due to the large broad-line region sizes and cosmological time dilation, which necessitates long-term monitoring programmes. Recent results on high-luminosity quasars \citep[e.g.][]{Lira_2018, Hoormann_2019, Grier_2019, Kaspi_2021} suggest a shallower radius-luminosity than measured in \citet{Mclure_2004}, which is more consistent with the photoionisation expectation of $R\propto L^{0.5}$. Black hole masses that are over-sensitive to the observed luminosity could obfuscate the underlying mass growth rates of the population, artificially boosting the apparent mass evolution when there is luminosity evolution.

Another way that the black hole masses can be overestimated is if they represent the tail end of the virial black hole mass error function. Because of the precipitous decline in abundance in the high mass regime, lower-mass black holes are more likely to be scattered high by a symmetric error function. The \citetalias{He_2024_H24} derivation of the intrinsic black hole mass function accounting for the error function estimated a high-mass slope of $\beta \sim -6$ as opposed to the $\beta \sim -3$ measured in Table \ref{tab:BHMF_model_params}. However, our quasars are selected by luminosity and not black hole mass, which is proportional to mass by $M_{\rm{BH}} \propto L^{b}$ in Equation \ref{eq:mgii_virial}, where $b = 0.5$ under standard photoionisation calculations and $b=0.62$ in the \citet{Shen_2011} calibration. This has a weaker effect on the black hole mass than the measured full-width at half maximum, to which the black hole mass is correlated by $M_{\rm{BH}} \propto {\rm{FWHM}}^{2}$, based on the virial theorem. Because the scatter in FWHM is not systematically biased by our luminosity selection, mass bias resulting from a fixed homoscedastic virial mass estimate error distribution would cancel out in a differential growth analysis between redshifts.

\subsubsection{Slower growth between $z=6-5$?} 
The $1/V_{\rm{max}}$ binned mass function measured with the E-XQR-30+ database disfavours scenarios where the mass growth between $z=6-5$ is more rapid than between $z=5-4$. Instead, the completeness corrected $z\sim6$ active black hole mass functions \citep{Shankar_2009, Willott_2010, Wu_2022_Demo} suggest that the evolution from $z=6-5$ could be even more gradual, which is counter to the general expectation that black holes in the early Universe need to evolve quickly to match observed masses at high-redshift. In principle, the slow growth of massive black holes from $z=6-5$ can be caused by cosmic downsizing, where the highest mass black holes are experiencing preferential mass starvation and turning off, but it is inconsistent with the more rapid growth seen between $z=5-4$, unless the black hole masses in \citetalias{He_2024_H24} are overestimated by uncorrected \civ\ virial estimates. 

The evolution of the quasar luminosity function is potentially enhanced between $z=6.0-5.5$ compared to $z=5.5-5.0$ \citep{Kashikawa_2015, Giallongo_2019, Grazian_2020, Santos_2021}. However, a constant evolution in the quasar luminosity function across the end of reionisation epoch between $z=6-5$ is not excluded. Based on the comparison with literature $z\sim6$ mass functions and the mass function extrapolated from the measured growth between $z=5-4$, our analysis indicates that the mean rate of change in the black hole mass function between $z=6-5$ is consistent with $z=5-4$, implying no change in the mass growth rate across $z=6-4$.

\subsubsection{Additional remarks}
New luminosity function analyses at high redshift focusing on bright quasars \citep{Onken_2022_QLF, Grazian_2022} updated previous determinations \citep[e.g.][]{Yang_2016, McGreer_2018, Niida_2020_N20} with an increased space density of bright quasars. This result was prefaced by surveys of the ultraluminous quasar population \citep[e.g.][]{Schindler_2017_ELQS, Schindler_2019, Boutsia_2020, Cristiani_2023_QUBRICS} which showed that previous surveys had underestimated their completeness corrections by $\sim30\%$ \citep{Schindler_2019_QLF, Boutsia_2021}. In the black hole mass function context, \citet{Wu_2022_Demo} attributed their high-mass divergence from \citet{Willott_2010} to their inclusion of an error model, which implies that the virial black hole masses of quasars within their highest mass bins are possibly overestimated. However, applying an upper redshift threshold of $z=6.3$ to \citet{Wu_2022_Demo}, as discussed in Section \ref{sec:overBHmass}, corrects their mass function to within $1\sigma$ of our model extrapolation from the observed $z=5-4$ $k_{\rm{ef}}$ growth. We expect that discrepancies between mass functions derived from heterogeneous methods to be an unavoidable consequence of the multitude of correction factors, broad scatter from uncertainties in the black hole mass and luminosity determinations, as well as cosmic variance. This highlights the importance of wide-area surveys of high completeness, as well as the need for improving the accuracy of black hole mass measurement techniques.

\subsection{Black hole initial mass function}
\begin{figure}
	\includegraphics[width=1.0\columnwidth]{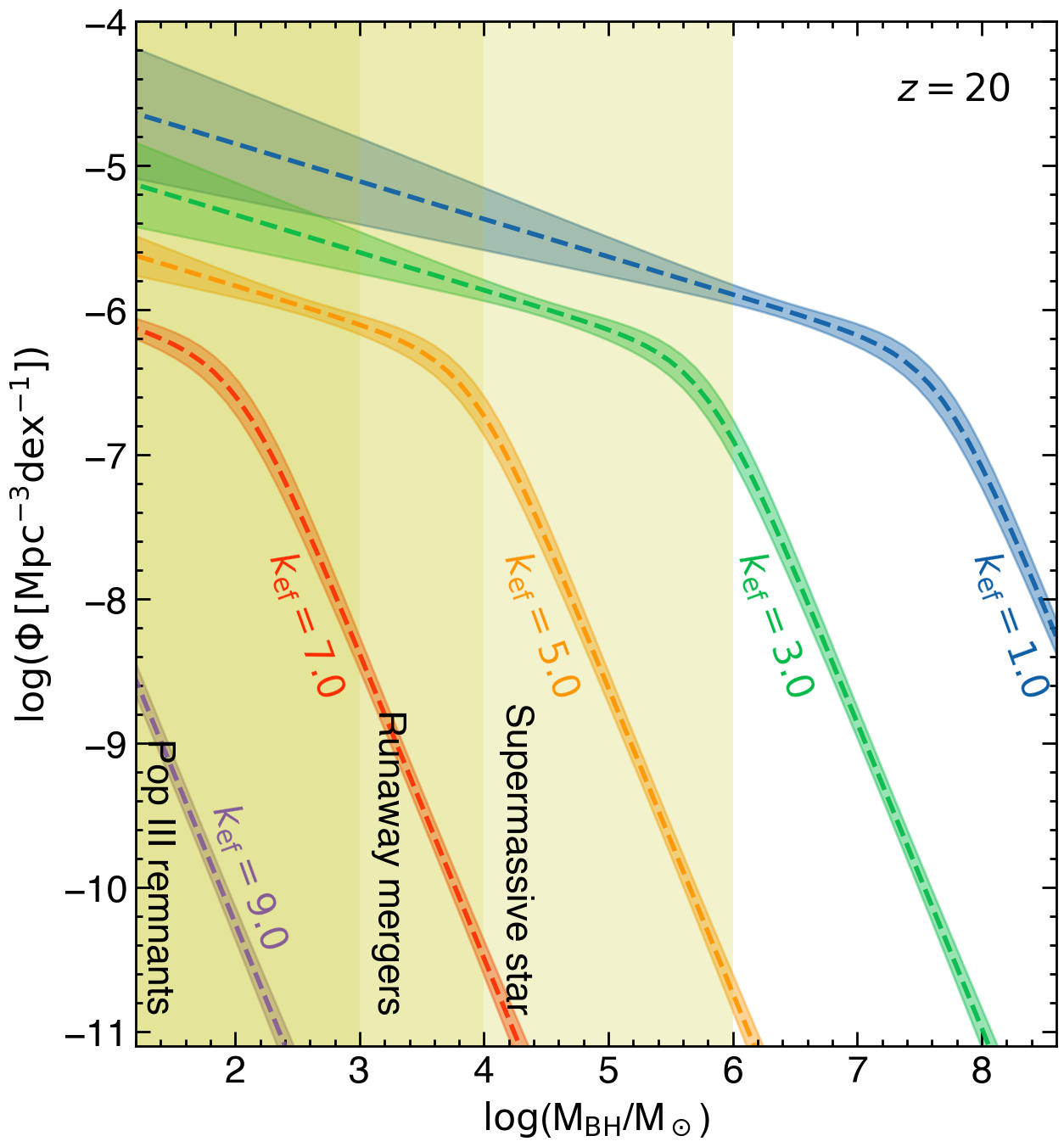}
    \caption[Black hole initial mass functions at $z=20$]{Black hole mass functions evolved to $z=20$ through the continuity equation based on the reference mass function model at $z\sim5$ and effective dimensionless growth parameters, $k_{\rm{ef}} = 1-9$, where $k_{\rm{ef}} = 9$ is the growth rate of a black hole growing 100\% of the time with an Eddington ratio of $\langle\lambda\rangle=1$ and radiative efficiency of $\epsilon=0.1$. Overplotted are three proposed seeding mechanisms for the most massive high-redshift black holes and their seed mass range: Pop III remnants $(< 10^3\,M_{\odot})$, runaway mergers $(10^3 - 10^4\, M_{\odot})$, and supermassive star collapse $(10^4 - 10^6\, M_{\odot})$ \citep[see review from][]{Inayoshi_2020}. A $z=20$ reference redshift for the initial black hole mass function corresponds to $\sim175$ Myr after the Big Bang.}
    \label{fig:seedz20}
\end{figure} 

Although there are a few supermassive $(>10^{10}\,M_{\odot})$ black holes in XQz5+, this dataset does not produce the most stringent constraints on black hole seeds due to their relatively low redshift compared to billion solar mass black holes observed at $z>7$ \citep[e.g.][]{Mortlock_2011, Banados_2018, Wang_2018, Yang_2019, Yang_2020_z7.5}. More recently, the James Webb Space Telescope enabled the search of AGN to push into higher redshifts \citep[e.g.][]{Larson_2023}, leading to the highest redshift AGN ($z\sim10$) discovery thus far, revealed by its X-ray emission as seen by the Chandra X-ray Observatory \citep{Bogdan_2024}. Its mass was inferred to be $10^7-10^8\,M_{\odot}$, comparable to the stellar mass of its host, which would require heavy seeds of $10^4-10^5\,M_{\odot}$ even with constant accretion at the Eddington limit.

Using the $z\sim5$ black hole mass function model derived in this study, we trace the evolution of the mass function back to $z=20$ ($\sim 175\, \rm{Myr}$ post Big Bang), which is an accessible redshift for the future Laser Interferometer Space Antenna \citep{Amaro-Seoane_2017}. We derive the hypothetical initial mass function at $z=20$ that would be consistent with the observed $z\sim5$ mass function for effective growth $k_{\rm{ef}} = 1-9$, where $k_{\rm{ef}} = 9$ is the growth rate of a black hole that is growing without stopping at the Eddington limit with the fiducial radiative efficiency of $\epsilon=0.1$. Figure \ref{fig:seedz20} presents the $z=20$ initial mass functions with proposed seeding mechanisms and their associated seed masses: Pop III remnants $(< 10^3\,M_{\odot})$, runaway mergers $(10^3 - 10^4\, M_{\odot})$, and supermassive star collapse $(10^4 - 10^6\, M_{\odot})$ \citep{Inayoshi_2020}. Black holes would need to grow at $k_{\rm{ef}} > 5.0$ since their formation at $z=20$ in order for the observed mass function to be consistent with the maximum mass of heavy seeds. Even the most massive black holes of XQz5+ are consistent with Pop III remnants if allowing for perpetual Eddington limited growth at the fiducial radiative efficiency of $\epsilon = 0.1$. Unlike supermassive black holes at redshifts $z>6$, the lower redshift black holes are not as strongly limited by the available cosmic time and thus there are multiple channels of mass assembly that are consistent with the observed spatial and mass densities. 

\begin{figure}
	\includegraphics[width=1.0\columnwidth]{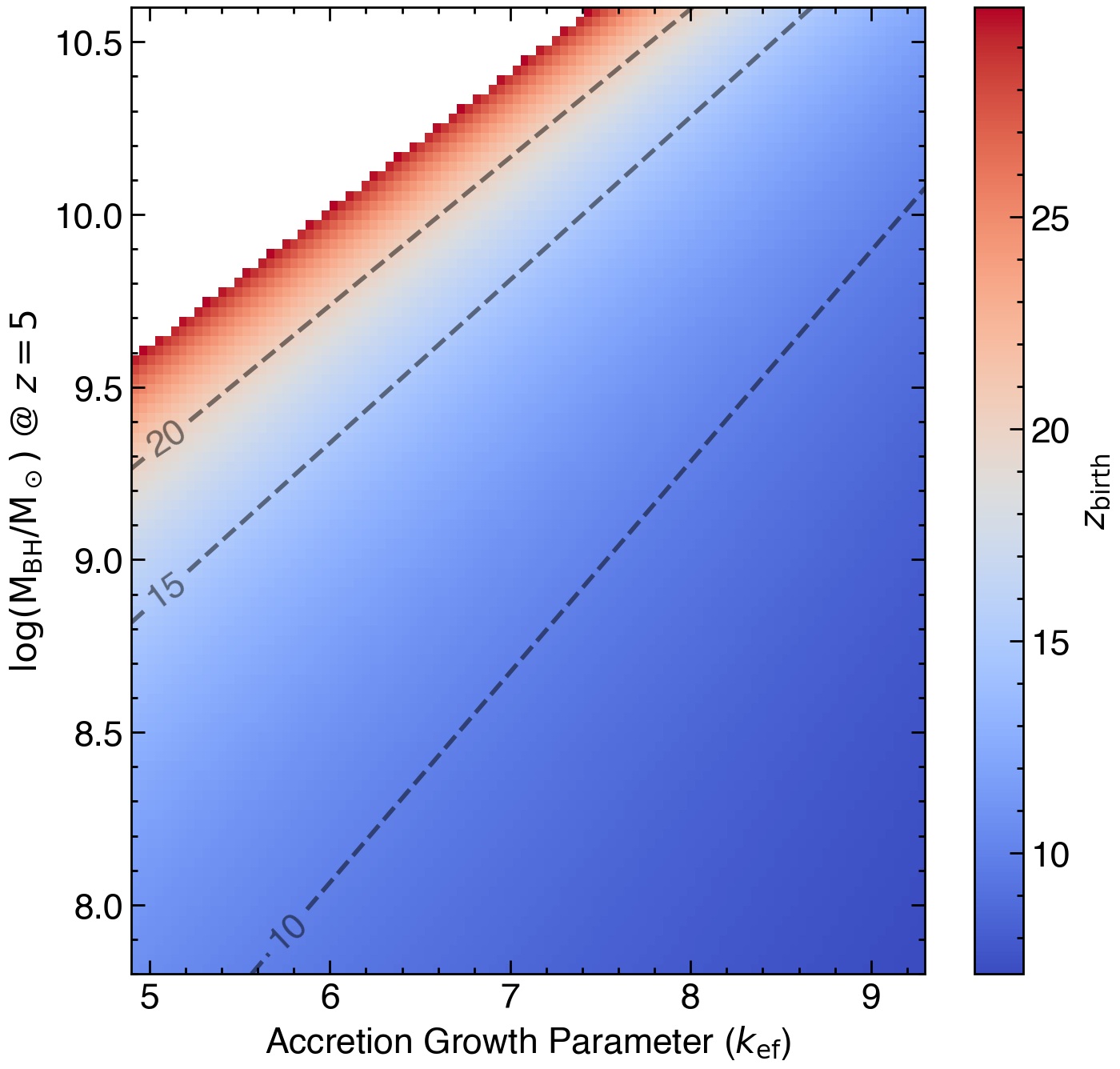}
    \caption[Growth rate of $10^4\,M_{\odot}$ for $z\sim5$ quasars]{Redshift of $10^4\,M_{\odot}$ black hole seed birth required to accrete a black hole of a specific mass by redshift $z=5$ for a variety of accretion growth rates. The white exclusion region is set by an upper redshift limit of $z=30$, which corresponds to an age of $\sim 100$ Myr. The ten billion solar mass black holes observed in XQz5+ are only consistent with $10^4\,M_{\odot}$ black hole seeds if the accretion growth rate is $\sim3.5$ times the observed growth between $z=5-4$ mass functions.}
    \label{fig:seedbirth}
\end{figure}

\begin{figure}
	\includegraphics[width=1.0\columnwidth]{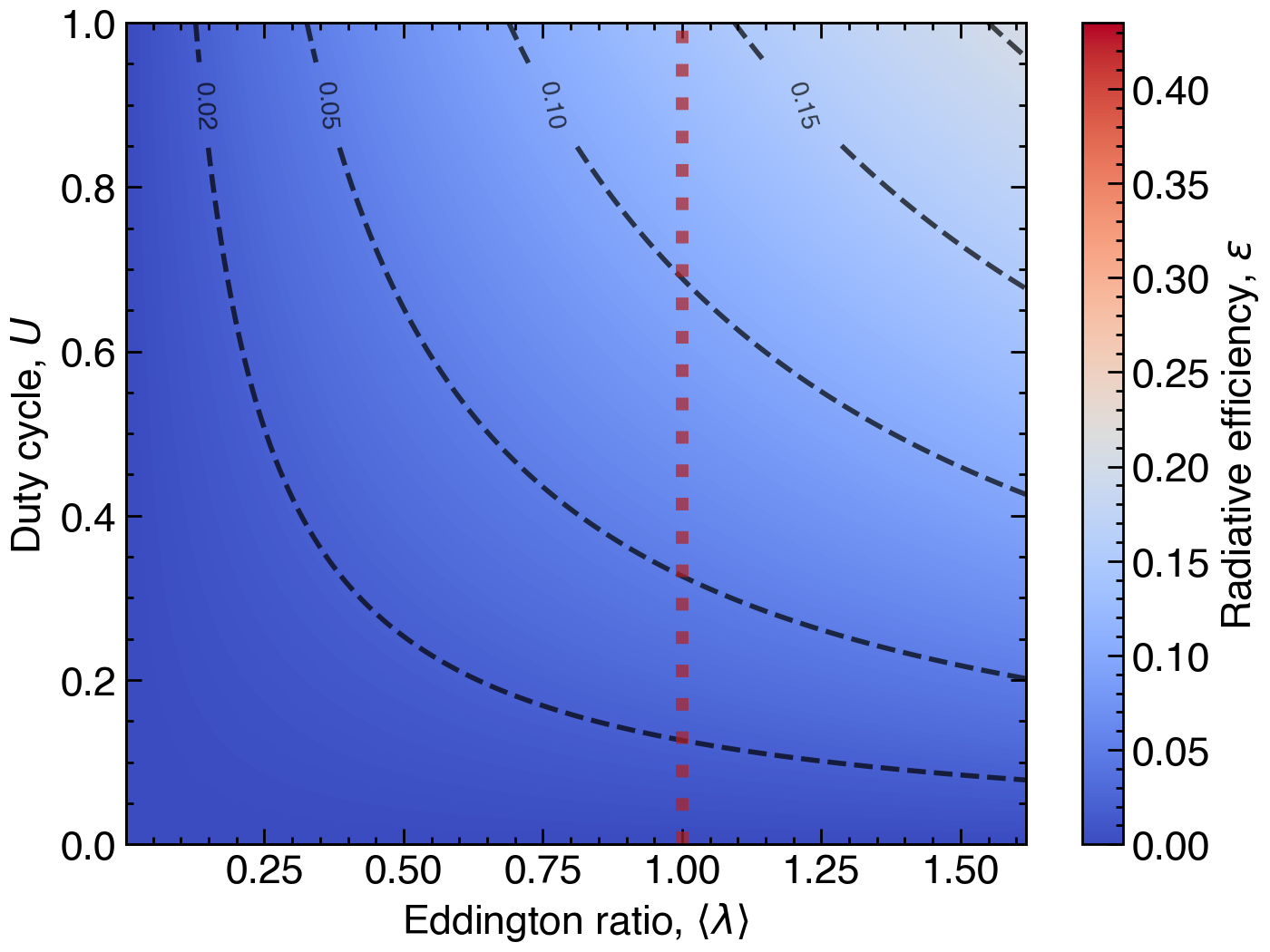}
    \caption[Joint constraint surface plot for $k_{\rm{ef}} = 6.2$]{Surface plot showing the relationship between the Eddington ratio, duty cycle, and radiative efficiency satisfying Equation \ref{eq:kef} for $k_{\rm{ef}} = 6.2$. We overplot dashed radiative efficiency contours and highlight $\langle\lambda\rangle = 1.0$ with the vertical dashed line.}
    \label{fig:surfacek6}
\end{figure}

In Figure \ref{fig:seedbirth}, we estimate the redshift at which a $10^4\,M_{\rm{\odot}}$ black hole would need to be born by to produce a black hole with $10^{10}\,M_{\odot}$ by $z=5$, given a particular growth rate, $k_{\rm{ef}}$. We choose $10^4\,M_{\rm{\odot}}$ as the fiducial seed mass because it is the interface between two of the heavy seed mechanisms: supermassive star collapse and runaway merger from a dense stellar cluster. We have set an upper limit to the birth redshift at $z=30$, which corresponds to $\sim 100$ Myr age for the universe. The results show that $10^4 \, M_{\odot}$ seeds could not produce the observed ten billion solar mass black holes by $z=5$ unless the mean growth rate exceeds $k_{\rm{ef}} = 6.2$. As shown in Figure \ref{fig:surfacek6}, the joint constraints on the radiative efficiency, Eddington ratio, and duty cycle from a growth rate of $k_{\rm{ef}} > 6$ would exclude $\epsilon > 0.14$ for Eddington limited accretion.

In Figure \ref{fig:seedbirth}, black holes with masses $\log{(\rm{M_{BH}}/\rm{M_\odot})} \gtrsim 9.5$ at $z\sim5$ are excluded by the upper redshift limit $z_{\rm{max}}=30$ when $k_{\rm{ef}} = 5$ and the seed mass is $10^4 \,M_{\odot}$. The majority of XQz5+ black holes are more massive than $\log{(\rm{M_{BH}}/\rm{M_\odot})} = 9.5$, which provides evidence that the rate of growth by accretion in the early universe is likely to be a factor of $\sim3.5$ higher than observed between $z=5-4$. If the seed mass is within the range of $10^{5-6}\,M_{\odot}$ instead, indicative of a direct collapse heavy black hole seed, then growth rates $\sim3$ times that of $z=5-4$ is sufficient to explain the $z\sim5$ mass function with $z_{\rm{birth}} = 30$ as shown in Figure \ref{fig:seedz20}. Otherwise, if faster evolution is not occurring at higher redshifts, a growth factor of $k_{\rm{ef}} = 1.79$ extrapolated to $z=30$ would require seeds of $\sim10^{8.2}\,M_{\odot}$ to produce $10^{10}\,M_{\odot}$ black holes by $z=5$.

\section{Summary and Conclusion} \label{sec:conclusion}

In this study, we analyse the black hole mass function at $z\sim5$ and compare to mass functions at $z\sim4$ and $z\sim6$. We measure the evolution in the mass function between $z=5-4$ using the continuity equation and present joint constraints on the duty cycle, radiative efficiency, and mean Eddington ratio. Finally, we discuss the implications of our study on the population of black hole seeds in the early universe. The main results of this study are summarised:

\begin{itemize}
    \item Our sample, collated from \citet{Lai_XQz5}, \citet{Trakhtenbrot_2011}, and \citet{Lopez_2016_XQ100}, collectively referred to as XQz5+, is composed of 72 of the most luminous quasars with spectroscopic follow-up observations over the redshift range $4.5 < z < 5.3$. The black hole masses of the quasars in the sample are measured using the single-epoch virial mass estimate based on the \mgii\ broad emission-line in their respective studies. We use the $1/V_{\rm{max}}$ approach (Eq. \ref{eq:binned-bhmf}) to derive population distribution functions and present our completeness corrections in Figure \ref{fig:completeness}. Our sample occupies a highly complete parameter space, enabling further demographic analysis.
    \item The Eddington ratio distribution function for XQz5+ (Fig. \ref{fig:ERDF}) is consistent with a log-normal function centered around $\langle\log\lambda\rangle = -0.21 \pm 0.03$, with dispersion $\sigma_{\lambda} = 0.30 \pm 0.02$ dex. In comparison to other observed Eddington ratio distributions at lower \citep{He_2024_H24} and higher \citep{Dodorico_2023_XQR30} redshifts, the Eddington ratio distribution of XQz5+ is centered on a higher value, but the dispersion is comparable. 
    \item The mass function of XQz5+ (Fig. \ref{fig:BHMF}) is consistent with a power-law on the high-mass regime with an artificial turnover at $\log{(\rm{M_{BH}}/\rm{M_\odot})} \lesssim 9.5$ caused by incompleteness. We model the mass function with a double power-law (Eq. \ref{eq:dbl_plaw}) using the low-mass constraints of \citet{He_2024_H24} translated from $z\sim4$ to $z\sim5$. The evolution from XQz5+ to the \citet{He_2024_H24} mass function implies an accretion growth factor of $k_{\rm{ef}} \equiv \langle\lambda\rangle U(1-\epsilon)/\epsilon = 1.79\pm0.06$, where $\lambda$ is the Eddington ratio, $U$ is the duty cycle, and $\epsilon$ is the radiative efficiency. Cosmic variance is the dominant source of error in the measured mass growth rate.
    \item We extrapolate the measured evolution of the black hole mass function from $z=5-4$ to $z\sim6$, comparing against the existing literature in Fig. \ref{fig:BHMFz6} and a high-redshift comparison sample composed of XQR-30 \citep{Dodorico_2023_XQR30} augmented by the \citet{Fan_2023_ARAA} compilation, which we call E-XQR-30+. Although the binned black hole mass function of E-XQR-30+ is qualitatively similar to our $z\sim6$ prediction, a detailed analysis is reserved for a future study due to heterogeneity in the sample construction and uncertainty in its completeness function. 
    \item We also estimate $k_{\rm{ef}} = 1.89\pm0.31$ within the XQz5+ sample  with an additional $0.5-0.8$ in systematic uncertainty due to cosmic variance. This result is consistent with the growth measured between $z=5-4$. 
    \item If the mean mass growth rate observed at $z\sim5$ is extended to an initial mass function at $z\sim20$, the seed mass required to form a $10^9\,M_{\odot}$ black hole by $z\sim5$ would be $>10^7\,M_{\odot}$. Therefore, we find that the mean mass evolution in the early universe would need to be a factor of $\gtrsim 3-4$ times the rate measured between $z\sim5-4$ to be consistent with heavy seed masses from exotic black hole formation mechanisms (Fig. \ref{fig:seedz20}).
\end{itemize}

From this study, we infer that the black hole mass growth has slowed considerably by $z\sim5$ and would have had to been more rapid in the early Universe. For accretion-dominated mass growth, this imposes more stringent constraints on the Eddington ratio, radiative efficiency, and duty cycle unless supercritical growth is achieved and maintained over extended durations. Improvements to these measurements will require measuring black hole masses for a larger sample of quasars over a wider area of the sky or expanding the survey flux limit. Improved statistics would allow for more advanced statistical modelling techniques to produce reliable joint constraints on distributions of black hole properties. Additionally, an enhanced understanding of the obscured quasar fraction across the relevant redshift and luminosity ranges would further refine our result. Future constraints on quasar demographics at higher redshift will depend on the results from new facilities such as the James Webb Space Telescope, which will enable black hole mass measurements from the \hbeta\ emission-line between redshifts $5 < z < 10$. The first homogeneous samples at such high redshifts will illuminate the circumstances that gave rise to the accelerated growth required for prospective supermassive black holes to reach their observed masses by $z\lesssim7$. The future Laser Interferometer Space Antenna has the capability for direct detections of black hole mergers with total participating masses of $\sim10^4-10^7$ beyond redshifts of $z\sim20$, probing the activity of newly-born massive seeds before they lose the memory of their birth. 

\section*{Acknowledgements}
We are grateful to the anonymous referee whose contribution has improved this study.

S.L. is grateful to the Research School of Astronomy \& Astrophysics at Australian National University for funding his Ph.D. studentship.

CAO was supported by the Australian Research Council (ARC) through Discovery Project DP190100252.

This paper is based on observations made with ESO Telescopes at the La Silla Paranal Observatory under programme IDs 084.A-0574(A), 084.A-0780(B), 087.A-0125(A), 094.A-0793(A), 098.A-0111(A), 0100.A-0243(A), 0104.A-0410(A), 108.22H9.001, 109.23D1.001, and 109.23D1.002.

The national facility capability for SkyMapper has been funded through ARC LIEF grant LE130100104 from the Australian Research Council, awarded to the University of Sydney, the Australian National University, Swinburne University of Technology, the University of Queensland, the University of Western Australia, the University of Melbourne, Curtin University of Technology, Monash University and the Australian Astronomical Observatory. SkyMapper is owned and operated by The Australian National University's Research School of Astronomy and Astrophysics. The survey data were processed and provided by the SkyMapper Team at ANU. The SkyMapper node of the All-Sky Virtual Observatory (ASVO) is hosted at the National Computational Infrastructure (NCI). Development and support of the SkyMapper node of the ASVO has been funded in part by Astronomy Australia Limited (AAL) and the Australian Government through the Commonwealth's Education Investment Fund (EIF) and National Collaborative Research Infrastructure Strategy (NCRIS), particularly the National eResearch Collaboration Tools and Resources (NeCTAR) and the Australian National Data Service Projects (ANDS).

The Pan-STARRS1 Surveys (PS1) and the PS1 public science archive have been made possible through contributions by the Institute for Astronomy, the University of Hawaii, the Pan-STARRS Project Office, the Max-Planck Society and its participating institutes, the Max Planck Institute for Astronomy, Heidelberg and the Max Planck Institute for Extraterrestrial Physics, Garching, The Johns Hopkins University, Durham University, the University of Edinburgh, the Queen's University Belfast, the Harvard-Smithsonian Center for Astrophysics, the Las Cumbres Observatory Global Telescope Network Incorporated, the National Central University of Taiwan, the Space Telescope Science Institute, the National Aeronautics and Space Administration under Grant No. NNX08AR22G issued through the Planetary Science Division of the NASA Science Mission Directorate, the National Science Foundation Grant No. AST-1238877, the University of Maryland, Eotvos Lorand University (ELTE), the Los Alamos National Laboratory, and the Gordon and Betty Moore Foundation.

The VISTA Hemisphere Survey data products served at Astro Data Lab are based on observations collected at the European Organisation for Astronomical Research in the Southern Hemisphere under ESO programme 179.A-2010, and/or data products created thereof.

This publication has made use of data from the VIKING survey from VISTA at the ESO Paranal Observatory, programme ID 179.A-2004. Data processing has been contributed by the VISTA Data Flow System at CASU, Cambridge and WFAU, Edinburgh.

This publication makes use of data products from the Two Micron All Sky Survey, which is a joint project of the University of Massachusetts and the Infrared Processing and Analysis Center/California Institute of Technology, funded by the National Aeronautics and Space Administration and the National Science Foundation.

This publication makes use of data products from the Wide-field Infrared Survey Explorer, which is a joint project of the University of California, Los Angeles, and the Jet Propulsion Laboratory/California Institute of Technology, and NEOWISE, which is a project of the Jet Propulsion Laboratory/California Institute of Technology. WISE and NEOWISE are funded by the National Aeronautics and Space Administration.

Software packages used in this study include \texttt{Numpy} \citep{Numpy_2011}, \texttt{Scipy} \citep{Scipy_2020}, \texttt{Astropy} \citep{Astropy_2013}, \texttt{PypeIt} \citep{Prochaska_2020_Pypeit}, \texttt{Specutils} \citep{specutils_2022}, \texttt{SpectRes} \citep{Carnall_2017}, and \texttt{Matplotlib} \citep{Matplotlib_2007}.

%%%%%%%%%%%%%%%%%%%%%%%%%%%%%%%%%%%%%%%%%%%%%%%%%%
\section*{Data Availability}
The data underlying this article will be shared on reasonable request to the corresponding author.



\bibliographystyle{mnras}
\bibliography{bibliography} % if your bibtex file is called example.bib

%\begin{thebibliography}{99}
%\end{thebibliography}

%%%%%%%%%%%%%%%%%%%%%%%%%%%%%%%%%%%%%%%%%%%%%%%%%%

%%%%%%%%%%%%%%%%% APPENDICES %%%%%%%%%%%%%%%%%%%%%

%\appendix
%\section{Additional Figures}

%%%%%%%%%%%%%%%%%%%%%%%%%%%%%%%%%%%%%%%%%%%%%%%%%%

% Don't change these lines
\bsp	% typesetting comment
\label{lastpage}
\end{document}